\documentclass[fleqn, usenatbib, useAMS]{mnras}

\usepackage{graphicx}
\usepackage{amsmath, amssymb}
\usepackage{newtxtext,newtxmath}
\usepackage[T1]{fontenc}
\usepackage{ae, aecompl}
\usepackage{xcolor}

\title[The geometry of radio emission from magnetic SPI]{Hunting for exoplanets via magnetic star-planet interactions: geometrical considerations for radio emission}

\author[Kavanagh \& Vedantham]{Robert D.~Kavanagh$^{1,2}$\thanks{Contact e-mail: \href{kavanagh@astron.nl}{kavanagh@astron.nl}} and Harish K.~Vedantham$^{1,3}$ \\ 
$^1$ASTRON, The Netherlands Institute for Radio Astronomy, Oude Hogeveensedijk 4, 7991PD, Dwingeloo, the Netherlands \\
$^2$Leiden Observatory, Leiden University, PO Box 9513, 2300 RA, Leiden, the Netherlands \\
$^3$Kapteyn Astronomical Institute, University of Groningen, Landleven 12, NL-9747AD Groningen, the Netherlands}

\date{Last updated ...; in original form ...}

\pubyear{2023}

\graphicspath{{figs/}}

\begin{document}
\label{firstpage}
\pagerange{\pageref{firstpage}--\pageref{lastpage}}
\maketitle


\begin{abstract}
Recent low-frequency radio observations suggest that some nearby M~dwarfs could be interacting magnetically with undetected close-in planets, powering the emission via the electron cyclotron maser (ECM) instability. Confirmation of such a scenario could reveal the presence of close-in planets around M~dwarfs, which are typically difficult to detect via other methods. ECM emission is beamed, and is generally only visible for brief windows depending on the underlying system geometry. Due to this, detection may be favoured at certain orbital phases, or from systems with specific geometric configurations. In this work, we develop a geometric model to explore these two ideas. Our model produces the visibility of the induced emission as a function of time, based on a set of key parameters that characterise magnetic star-planet interactions. Utilising our model, we find that the orbital phases where emission appears are highly dependent on the underlying parameters, and does not generally appear at the quadrature points in the orbit as is seen for the Jupiter-Io interaction. Then using non-informative priors on the system geometry, we show that untargeted radio surveys are biased towards detecting emission from systems with planets in near face-on orbits. While transiting exoplanets are still likely to be detectable, they are less likely to be seen than those in near face-on orbits. Our forward model serves to be a powerful tool for both interpreting and appropriately scheduling radio observations of exoplanetary systems, as well as inverting the system geometry from observations.
\end{abstract}



\section{Introduction}

The majority of exoplanets discovered to date orbit around low-mass main-sequence stars\footnote{\texttt{https://exoplanetarchive.ipac.caltech.edu}}, in agreement with formation theory \citep[][]{nicholson19, burn21}. M~dwarfs, the lowest mass stars on the main sequence, are the most numerous in the stellar neighbourhood \citep{winters19}, and are expected to preferentially host close-in rocky planets \citep{burn21, schlecker22}. While in theory the detection of an Earth-like planet orbiting an M~dwarf is much easier compared to a Sun-like star due to the higher mass/size ratio, these stars generally exhibit much higher levels of magnetic activity. As a result, the majority of these planets likely remain undetected to date via traditional techniques such as the radial velocity and transit methods, as the activity of the host star can readily drown out signatures of the planet.

That being said, an alternative mechanism may produce signatures which can be distinguished from stellar activity, particularly for M~dwarfs. This mechanism is thought to occur via magnetic star-planet interactions \citep[SPI;][]{zarka07, saur13}. The inspiration for this comes from Jupiter's sub-Alfv\'enic interactions with the Galilean moons Io, Europa, and Ganymede. The motion of these bodies through Jupiter's magnetosphere is known to produce bright coherent radio emission along the magnetic field line linking each moon to Jupiter, especially in the case of Io. The radio emission is powered by the electromotive force felt by charges in the ionospheres of the moons as they move across the Jovian magnetic field. This energy is transported towards Jupiter in the form of Alfv\'en waves \citep{alfven42, neubauer80}, which subsequently accelerate electrons that emit radio waves via the electron cyclotron maser (ECM) instability \citep{dulk85, treumann06}.

Determining if the orbit of a satellite is sub-Alfv\'enic or not requires knowledge of the plasma environment. In this region, the magnetic energy of the plasma exceeds the kinetic energy. Another way to express this is via the Alfv\'enic Mach number, which is
\begin{equation}
M_\textrm{A} = \frac{\Delta u}{u_\textrm{A}} = \frac{\Delta u \sqrt{4\pi\rho}}{B} ,
\end{equation}
where $\Delta u$ is the plasma velocity in the rest frame of the satellite, $u_\textrm{A}$ is the Alfv\'en velocity, and $\rho$ and $B$ are the density and magnetic field strength of the plasma at the position of the satellite. When the ratio of the velocities is less than unity ($M_\textrm{A} < 1$), the disturbance in the magnetic field created by the satellite can propagate as Alfv\'en waves along the field lines back to the star. If $M_\textrm{A} > 1$ however, the disturbance created by the satellite is moving faster than the Alfv\'en waves and therefore a shock discontinuity is set up and the disturbance can no longer flow back to the star. The boundary where $M_\textrm{A} = 1$ is known as the Alfv\'en surface, which can be complex in shape depending on the magnetic field topology at the stellar surface \citep{vidotto14}.

The reason why M~dwarfs are excellent candidates for the same type of interactions as seen with Jupiter and its inner moons is primarily due to the strong magnetic fields they possess, which can be upwards of a kilogauss (kG) in strength \citep{kochukhov21}. High field strengths correspond to high Alfv\'en velocities, meaning that the plasma, or wind in the case of a low-mass main-sequence star, must be accelerated to high velocities before $M_\textrm{A} > 1$. As a result, M~dwarfs are likely to harbour large Alfv\'en surfaces, enclosing a wide range of orbits wherein magnetic SPI can occur \citep[see][]{kavanagh21, kavanagh22}.

There has been a resurgence in the search for magnetic SPI in recent years, primarily due to the detection of bright radio emission with a high degree of circular polarisation from nearby M~dwarfs \citep{vedantham20, perez-torres21, callingham21b, pineda23, trigilio23}, which is a signpost of the ECM mechanism \citep{dulk85}, although not necessarily powered by SPI. In the case of the 19 M~dwarfs detected by \citet{callingham21b}, none show any correlation between their radio luminosities and activity indicators. This is consistent with the driving mechanism being magnetospheric in origin. Yet, none of these stars are known to host close-in planets, leaving the interpretation ambiguous \citep[see however the recent detection by][]{blanco-pozo23}. If the detected emission from these systems is in fact due to the presence of undiscovered companions, there is the question of is there something special about these systems? If so, what is it about these systems that makes them more visible compared to other nearby M~dwarfs?

ECM emission is beamed, and is generally only visible for brief windows. A result of this can be seen from the emission Io induces on Jupiter, which appears only at `quadrature' points of Io's orbit (orbital phases of 0.25 and 0.75). To determine precisely when emission will appear for a system requires both knowledge of the geometry of the large-scale magnetic field that the satellite interacts with, as well as the properties of the emission cone generated from the interaction \citep{kavanagh22}. It could be the case that certain combinations of the geometry of the stellar magnetic field and planetary orbit could produce emission that is more visible compared to other configurations. We note also that the planet itself could be a source of beamed radio emission \citep{ashtari22}, which could be difficult to disentangle from the emission induced on the star. However, there are many uncertainties in the frequency at which we expect exoplanetary radio emission, primarily due to our lack of knowledge about exoplanetary magnetic fields \citep[see also][]{kavanagh19}.

Recently, we utilised magnetohydrodynamic (MHD) models to assess the beaming of emission induced by a hypothetical planet for a variety of orbits around WX~UMa \citep{kavanagh22}, one of the M~dwarfs detected by \citet{callingham21b}. The method used was based on the surface magnetic field map of the star obtained using the Zeeman-Doppler imaging technique \citep[ZDI; see][]{donati09, kochukhov21}. However, these maps are not generally available for M~dwarfs. In fact, the only other star in the sample presented by \citet{callingham21b} with a magnetic field map is AD~Leo \citep{morin08, lavail18}. Note that \citet{callingham21b} suggest the detected emission could be in fact due to flaring, and not magnetic SPI. 

Our work on WX~UMa illustrated that sophisticated MHD models can help us to better understand the underlying mechanism generating ECM emission on nearby M~dwarfs, particularly in terms of identifying potential signatures of undiscovered planets. However, they are reliant on the availability of magnetic field maps for M~dwarfs, and are also computationally expensive. Therefore, there is a mounting need for an alternative method to estimate the visibility of planet-induced radio emission that does not heavily depend on ZDI and MHD simulations. This would allow for the detected emission reported by \citet{callingham21b}, as well as future observations, to be better-interpreted. The Exoplanetary and Planetary Radio Emission Simulator (ExPRES) code developed by \citet{hess11} is suitable for this in theory, which was originally developed to model the observed auroral emission on Jupiter and Saturn. To our knowledge however, it has not been utilised to answer the questions laid out in this work. We discuss the comparison between our methods in this work to the ExPRES code in Section~\ref{sec:expres discusson}.

In this paper, our main goal is to answer two questions:
\begin{enumerate}
\item What orbital phases is radio emission most likely to appear at in magnetic SPI?
\item What systems are we more likely to detect in untargeted radio surveys?
\end{enumerate}
To answer these questions, and also address the issues mentioned above, we develop a forward model based on key parameters relating to the geometry of magnetic SPI to predict the visbility of planet-induced radio emission as a function of time. This model provides the community with a flexible tool to interpret radio observations from low-mass stars in the context of magnetic SPI. The model is described in Section~\ref{sec:model}. In Section~\ref{sec:orbital phases}, we illustrate the use of the model by demonstrating the phenomenon of emission appearing at quadrature points of a satellite's orbit, as is seen for Jupiter's moon Io. Then in Section~\ref{sec:bias}, we utilise the model to address the question of are we systematically biased towards detecting emission from systems with certain architectures.


\section{\textsc{MASER}: A code for modelling magnetic star-planet interactions}
\label{sec:model}

In this Section, we describe the model we develop to predict when radio emission induced on a star via magnetic SPI is visible as a function of time. The model is freely available as a Python code on GitHub as the \textsc{MASER} (Magnetically interActing Stars and Exoplanets in the Radio) code\footnote{\texttt{github.com/robkavanagh/maser}}. The model takes a key set of inputs relating to the geometry and physical properties of magnetic SPI, as well as an array of times for which the visibility of the radio emission is computed. Table~\ref{tab:model parameters} lists each quantity and their respective symbols, which we use throughout unless noted otherwise.

The MASER code computes what we refer to as the `visibility lightcurve' for the system described by the input parameters (described further in Section~\ref{sec:aligned misaligned}). The code depends only on NumPy \citep{numpy}. It is also compatible with Numba \citep{numba}, which allows for quick execution. When utilised with Numba's `no Python mode', a lightcurve with $10^4$ time elements takes 2.5 milliseconds to compute on average using a single performance core of an Apple M2 chip, which is about 50 times faster than the standard computation time using Python.

\begin{table}
\caption{Reference list for the parameters of the \textsc{MASER} code.}
\label{tab:model parameters}
\centering
\begin{tabular}{lc}
\hline
Parameter & Symbol \\
\hline
\underline{Star}: \\
Mass & $M_\star$ \\
Radius & $R_\star$ \\
Rotation period & $P_\star$ \\
Inclination & $i_\star$ \\
Rotation phase at time zero & $\phi_{\star,0}$ \\
Dipolar field strength at the pole & $B_\star$ \\
Magnetic obliquity & $\beta$ \\
\underline{Planet}: \\
Orbital distance & $a$ \\
Orbital inclination & $i_\textrm{p}$ \\
Projected spin-orbit angle & $\lambda$ \\
Orbital phase at time zero & $\phi_{\textrm{p},0}$ \\
\underline{Radio emission}: \\
Observing frequency & $\nu$ \\
Cone opening angle & $\alpha$ \\
Cone thickness & $\Delta\alpha$ \\
\hline
\end{tabular}
\end{table}


\subsection{The geometry of magnetic star-planet interactions}
\label{sec:geometry}

To determine if radio emission induced on stars by an orbiting planet is visible at a given time, we first need to establish the key physical and geometrical parameters of the system. The host star has a mass $M_\star$, radius $R_\star$, and rotation period $P_\star$. Its rotation axis is $\hat{z}_\star$, which is inclined relative to the line of sight $\hat{x}$ by the angle $i_\star$. The star rotates about $\hat{z}_\star$ in a clockwise direction when looking along $\hat{z}_\star$. Note that all vectors denoted with a hat are unit vectors (their magnitude is unity). 

Given that our focus here is on M~dwarfs, we opt to represent the large-scale magnetic field of the star that the planet interacts with as a dipole. Dipolar magnetic fields drop off in strength slowest as a function of distance $r$ compared to higher order modes (quadrupole, octupole, etc.), with the field strength going as $r^{-3}$. As a result, unless the planet is very close to its host star, the field that the planet sees is a dipole. In addition to this, M~dwarfs often exhibit strong, predominantly-dipolar magnetic fields \citep[see][]{donati08, morin08, morin10}.

The maximum magnetic field strength at the stellar surface is $B_\star$, which for a dipolar field occurs at its magnetic poles. The magnetic axis of the star $\hat{z}_B$ points outward from the center of the star to the Northern magnetic pole, and is tilted relative to the stellar rotation axis $\hat{z}_\star$ by the angle $\beta$. This is known as the magnetic obliquity. When $\beta \neq 0$, the magnetic axis precesses about the stellar rotation axis as the star rotates. We assume that the magnetic field rotates rigidly with the stellar rotation period. At time $t$, the rotation phase of the star is
\begin{equation}
\phi_\star = \phi_{\star,0} + \frac{t}{P_\star},
\end{equation}
where $\phi_{\star,0}$ is the stellar rotation phase at $t = 0$. Note that from this definition, the phase varies from 0 to 1. As such, we multiply the phase by $2\pi$ when used in trigonometric functions. In Appendix~\ref{sec:star vectors}, we describe the coordinate system for the stellar rotation and magnetic field in more detail.

\begin{figure}
\centering
\includegraphics[width = \columnwidth]{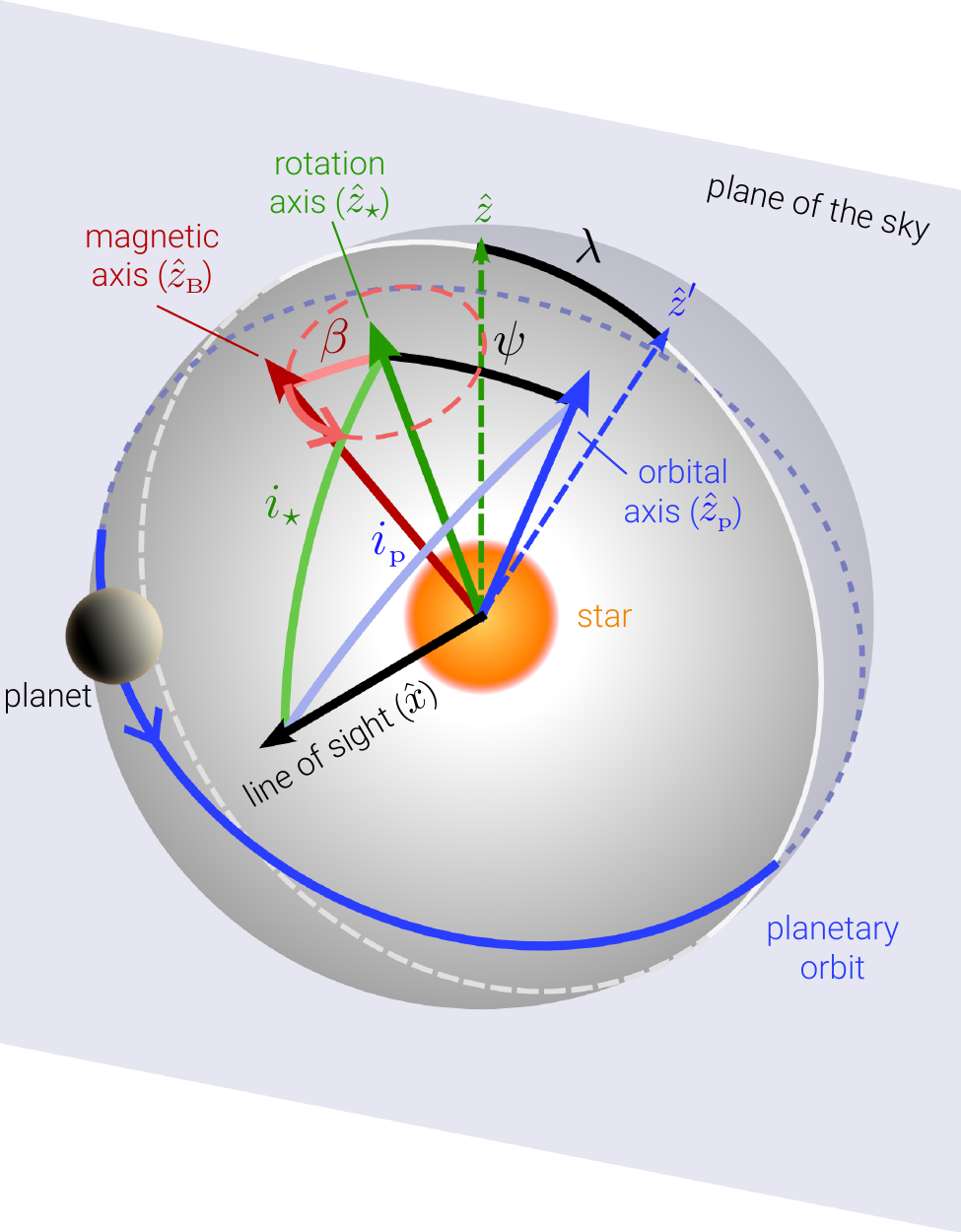}
\caption{A sketch illustrating the geometric properties relevant to interactions between a planet and the large-scale magnetic field of its host star. The star spins around its rotation axis $\hat{z}_\star$ in a clockwise direction when looking along $\hat{z}_\star$ (i.e. in a right-handed configuration). The rotation axis is inclined relative to the line of sight $\hat{x}$ by the angle $i_\star$. The star also possesses a large-scale dipolar magnetic field, with the magnetic axis $\hat{z}_\textrm{B}$ representing the position of the Northern magnetic pole. The magnetic axis is tilted relative to the rotation axis by the angle $\beta$, which is known as the magnetic obliquity. When $\beta \neq 0\degr$ or $180\degr$, the magnetic field precesses about the rotation axis, as indicated by the dashed red line. The planet is in a circular orbit around the star, orbiting with a right-handed configuration about its orbital axis $\hat{z}_\textrm{p}$. Its orbital axis is inclined relative to the line of sight by the angle $i_\textrm{p}$. In general, the line of sight, rotation, and orbital axes do not lie in the same plane due to spin-orbit misalignment. This is described by the angle $\psi$ that is formed between the rotation and orbital axes. The projection of the rotation and orbital axes onto the plane of the sky ($\hat{z}$ and $\hat{z}'$ respectively) form the angle $\lambda$, which is known as the \textit{projected} spin-orbit angle.}
\label{fig:sketch geometry}
\end{figure}

Around the star, a planet orbits at a distance $a$. Its orbital period $P_\textrm{p}$ is provided via Kepler's third law: 
\begin{equation}
P_\textrm{p} = 2\pi\sqrt{\frac{a^3}{GM_\star}} ,
\end{equation}
where $G$ is the gravitational constant. We assume that the planet's orbit is circular. Its position is described by the vector $\hat{x}_\textrm{p}$, and the vector normal to its orbital plane $\hat{z}_\textrm{p}$ is inclined relative to $\hat{x}$ by the angle $i_\textrm{p}$. Again, the convention we adopt for the orbit direction is clockwise when looking along $\hat{z}_\textrm{p}$. The vector $\hat{z}_\textrm{p}$ is misaligned with respect to $\hat{z}_\star$ by the angle $\psi$, known as the spin-orbit angle. Note that in general, it is easier to measure the \textit{projected} spin-orbit angle $\lambda$ for exoplanetary systems, which is the angle between $\hat{z}$ and $\hat{z}'$, the projections of $\hat{z}_\star$ and $\hat{z}_\textrm{p}$ on to the plane of the sky \citep{triaud18}. The relation between $\psi$ and $\lambda$ is given by Equation~\ref{eq:spin orbit angle}. The orbital phase of the planet at time $t$ is 
\begin{equation}
\phi_\textrm{p} = \phi_{\textrm{p},0} + \frac{t}{P_\textrm{p}},
\end{equation}
where $\phi_{\textrm{p},0}$ is the orbital phase at $t = 0$. Again, the values for $\phi_\textrm{p}$ range from 0 to 1. When $\phi_\textrm{p} = 0$, the planet is closest to the observer (at conjunction). However, if $i_\textrm{p} = 0\degr$ or $180\degr$, the planet is always at the same distance from the observer, and the planet's position is either in the direction of $-\hat{z}'$ or $\hat{z}'$ respectively at $\phi_\textrm{p} = 0$. Appendix~\ref{sec:planet vectors} presents the details of the coordinates for the planet and spin-orbit misalignment. A geometric sketch of the quantities introduced here is shown in Figure~\ref{fig:sketch geometry}.

\subsection{Interactions with dipolar magnetic fields}
\label{sec:dipole}

With the relevant properties of the exoplanetary system established, we now describe the magnetic field of the star in more detail. The shape of a dipolar magnetic field line is described by the following equation \citep{kivelson95}:
\begin{equation}
r = L\sin^2\theta .
\label{eq:dipolar field radius}
\end{equation}
Here, $r$ is the radius of a point on the field line measured from the center of the star, $\theta$ is the magnetic co-latitude of the point, which is measured from the direction that $\hat{z}_\textrm{B}$ points in, and $L$ is the distance between the center of the star and the magnetic field line at the magnetic equator.

\begin{figure}
\centering
\includegraphics[width = \columnwidth]{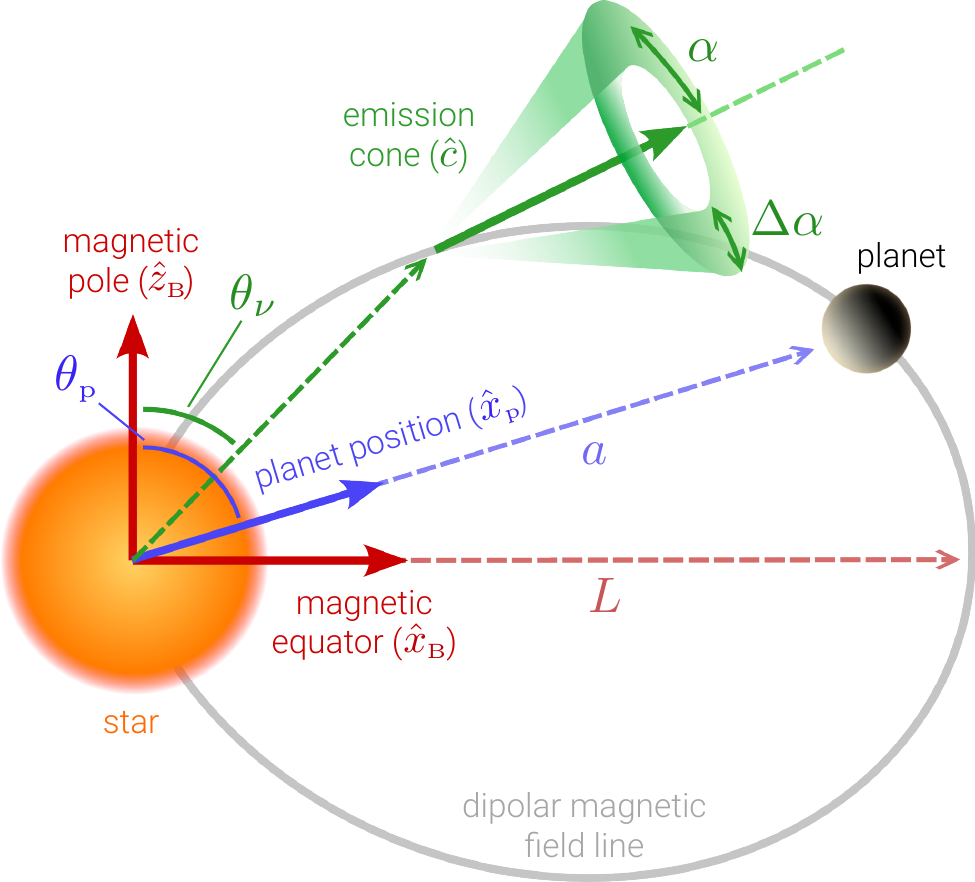}
\caption{Sketch of a planet interacting with a dipolar magnetic field line of a star of size $L$. The planet orbits at a distance $a$ from the star, and its position is described by the vector $\hat{x}_\textrm{p}$. It forms the angle $\theta_\textrm{p}$ with the magnetic axis of the star $\hat{z}_\textrm{B}$. Assuming the interaction occurs sub-Alfv\'enically, the planet induces the generation of radio emission along the line connecting it to the star at a distance $r_\nu$ and magnetic co-latitude $\theta_\nu$. This emission is beamed, and propagates outward in a hollow cone. The cone has a characteristic opening angle $\alpha$ and thickness $\Delta\alpha$, and is aligned with the vector $\hat{c}$, which in the Northern magnetic hemisphere is $\hat{c} = \vec{B} / B$, the normalised magnetic field vector at the emitting point. For clarity, we only show the emission cone in the Northern magnetic hemisphere. Note that in the Southern hemisphere, emission cones are aligned with the vector $\hat{c} = - \vec{B} / B$.}
\label{fig:sketch field line}
\end{figure}

At each point in the planet's orbit, it interacts with a field line of size $L$, which has a certain orientation relative to the line of sight. A sketch of this is shown in Figure~\ref{fig:sketch field line}. The magnetic co-latitude of the planet $\theta_\textrm{p}$ at a given time is determined by both its position and the direction of the magnetic axis:
\begin{equation}
\cos\theta_\textrm{p} = \hat{z}_\textrm{B} \cdot \hat{x}_\textrm{p} .
\label{eq:planet magnetic co-latitude}
\end{equation} 
With an orbital distance of $a$, Equation~\ref{eq:dipolar field radius} can then be rewritten as an expression for the size of the field line the planet interacts with at each point in its orbit:
\begin{equation}
L = \frac{a}{\sin^2\theta_\textrm{p}} .
\label{eq:field line size}
\end{equation}

To determine the orientation of the field line relative to the observer, we require the vector $\hat{x}_\textrm{B}$, which points along the magnetic equator of the field line that the planet interacts with. The planet's position can be expressed in terms of this vector along with $\hat{z}_\textrm{B}$ (see Figure~\ref{fig:sketch field line}):
\begin{equation}
\hat{x}_\textrm{p} = \sin\theta_\textrm{p}\hat{x}_\textrm{B} + \cos\theta_\textrm{p}\hat{z}_\textrm{B} .
\end{equation}
Re-arranging, $\hat{x}_\textrm{B}$ is:
\begin{equation}
\hat{x}_\textrm{B} = \frac{\hat{x}_\textrm{p}}{\sin\theta_\textrm{p}} - \frac{\hat{z}_\textrm{B}}{\tan\theta_\textrm{p}} .
\end{equation}
Knowing the directions of $\hat{z}_\textrm{B}$ and $\hat{x}_\textrm{B}$ as a function of time provides us with the direction of the emission cone $\hat{c}$ on the field line, which in turn determines if the radio emission the planet induces along the field line via sub-Alfv\'enic interactions is detectable (see Section~\ref{sec:radio}).

There is a caveat in assuming purely dipolar magnetic field lines for the star. Following from Equation~\ref{eq:field line size}, $L$ becomes very large for small values of $\theta_\textrm{p}$. However, it is not realistic for the star to have closed field lines that extend to hundreds of stellar radii, as the wind of the star will tend to blow them open once the kinetic wind energy exceeds the magnetic tension of the field line. Therefore, we adopt a maximum size for the field lines of 100~$R_\star$. If the size of the field line exceeds this, we limit the interaction to the hemisphere the planet is in only. In other words, if $L > 100~R_\star$ and $\theta_\textrm{p} < \pi/2$, the planet induces emission in the Northern magnetic hemisphere only, and if $L > 100~R_\star$ and $\theta_\textrm{p} > \pi/2$, it induces emission in the Southern magnetic hemisphere only. For sufficiently small orbits/large magnetic co-latitudes however, the planet orbits in the closed-field region of the star's magnetosphere. In this scenario, the planet induces emission in both magnetic hemispheres of the star \citep[e.g.][]{kavanagh21, kavanagh22}, similar to what is observed for the Io-induced radio emission on Jupiter \citep{marques17}.

\subsection{Radio emission from magnetic star-planet interactions}
\label{sec:radio}

When a conducting body moves through a magnetised plasma with a sub-Alfv\'enic velocity, mechanical waves known as Alfv\'en waves are produced \citep{alfven42, drell65}. In a planetary context, these waves are thought to travel along magnetic field lines, accelerating electrons in the process. Electrons accelerated with sufficiently large pitch angles (the angle between the velocity and local magnetic field vectors) are thought to experience a magnetic mirroring effect. The mirrored electrons have a so-called `loss-cone' velocity distribution, which are unstable to electromagnetic waves at the local cyclotron frequency \citep{treumann06}. Due to this instability, these electrons release their energy as electromagnetic waves via the electron cyclotron maser (ECM) instability, typically in the radio regime \citep{treumann06}. ECM emission occurs at the fundamental and harmonics of the local cyclotron frequency \citep{dulk85}, which in CGS units is
\begin{equation}
\nu_\textrm{c} = 2.8 B~\textrm{MHz}, 
\label{eq:cyclotron frequency}
\end{equation}
where $B$ is in Gauss (G).

\begin{figure}
\centering
\includegraphics[width = \columnwidth]{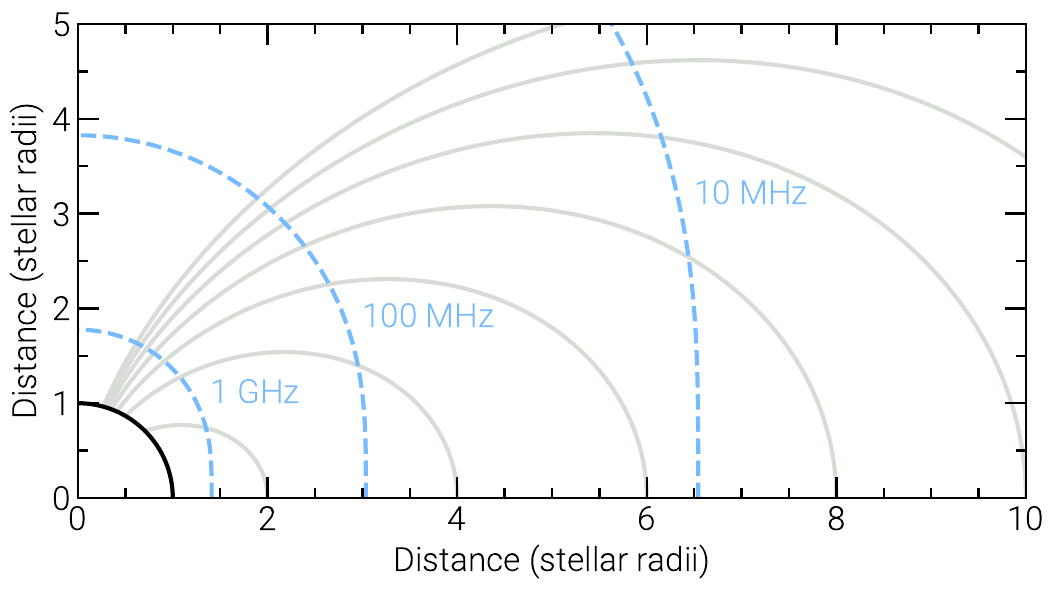}
\caption{Illustration of the large-scale dipolar magnetic field of a star. The $x$ and $y$ axes are aligned with the magnetic equator and pole respectively. The grey lines show field of various sizes, and the dashed blue lines illustrate the regions corresponding to ECM emission at 10~MHz, 100~MHz, and 1~GHz. The star is shown in the bottom left, which has a polar magnetic field strength of 1 kG.}
\label{fig:dipole lines}
\end{figure}

Equation~\ref{eq:cyclotron frequency} tells us that the emission frequency is a direct probe of the magnetic field strength at which the emission is generated. The field strength at each point on a dipolar field line, which is described by Equation~\ref{eq:dipolar field radius}, is given by \citep{kivelson95}:
\begin{equation}
B = \frac{B_\star}{2} \Big( \frac{R_\star}{r} \Big)^3 (1 + 3 \cos^2 \theta)^{1/2} .
\label{eq:field line strength}
\end{equation}
Using Equation~\ref{eq:dipolar field radius}, we can rewrite Equation~\ref{eq:field line strength} in terms of $r$ only, giving
\begin{equation}
B = B_\star \Big(\frac{R_\star}{r} \Big)^3 \Big(1 - \frac{3r}{4L}\Big)^{1/2}.
\label{eq:field line strength - radius}
\end{equation}
An example of the shape of dipolar field lines of different sizes along with corresponding regions of different cyclotron frequencies is shown in Figure~\ref{fig:dipole lines}.

As mentioned in the previous Section, we allow emission to be generated in both magnetic hemispheres if the size of the field line $L$ is less than 100~$R_\star$. To determine if fundamental ECM emission generated along the star-planet field line in either hemisphere at the frequency $\nu$ is visible to the observer, we first need to find the radius $r_\nu$ and magnetic co-latitude $\theta_\nu$ on the line that give a field strength $B_\nu = \nu / 2.8$ via Equation~\ref{eq:field line strength - radius}. As the field line is symmetric about the magnetic equator, the frequency at the point $(r_\nu,~\theta_\nu)$ is equivalent to that at $(r_\nu,\pi-\theta_\nu)$. Setting Equation~\ref{eq:field line strength - radius} equal to $B_\nu$ and re-arranging, we can define a new parameter $F$, which goes to zero as $r$ approaches $r_\nu$:
\begin{equation}
F = \Big(\frac{B_\nu}{B_\star}\Big)^2 \Big(\frac{r}{R_\star}\Big)^6 + \frac{3r}{4L} - 1.
\label{EQ:DIPOLAR FIELD LINE SOLVE}
\end{equation}
To the best of our knowledge, there is no analytical solution to $F=0$. Therefore, we utilise Newton's method find its root (see Appendix~\ref{sec:newthon method} for details).

Once we find $r_\nu$, we obtain $\theta_\nu$ via Equation~\ref{eq:dipolar field radius}. The point $(r_\nu,~\theta_\nu)$ corresponds to the Northern hemisphere, and $(r_\nu,\pi - \theta_\nu)$ corresponds to the Southern hemisphere. We then determine the direction of the magnetic field vector at the emitting point $\vec{B}_\nu$ in each magnetic hemisphere (see Appendix~\ref{sec:field line vector}), which in turn tells us the direction of the emission cone for each hemisphere $\hat{c}$. In the Northern magnetic hemisphere, the emission cone is parallel with the magnetic field vector ($\hat{c} = \vec{B}_\nu / B_\nu$), and in the Southern magnetic hemisphere, it is anti-parallel ($\hat{c} = -\vec{B}_\nu / B_\nu$). The angle between the line of sight $\hat{x}$ and the vector $\hat{c}$ determines if the radio emission is beamed towards the observer \citep{kavanagh22}. This angle is
\begin{equation}
\cos\gamma = \hat{x}\cdot\hat{c} .
\label{eq:beam angle}
\end{equation}
Note that emission from each hemisphere will have opposite circular polarisations, under the assumption that the magnetoionic mode is the same \citep{das21}.

The emission cone has a characteristic opening angle $\alpha$, and thickness $\Delta\alpha$. When $\gamma$ is in the range of $\alpha\pm\Delta\alpha/2$, the emission is visible to the observer (see Figure~\ref{fig:sketch field line}). According to \citet{melrose82}, the cone opening angle and thickness depend on the velocity of the accelerated electrons $u$, such that $\cos\alpha = u/c$ and $\Delta\alpha = u/c$~rad, where $c$ is the speed of light. In other words, $\alpha$ cannot exceed $90\degr$. For the Io-induced emission on Jupiter, opening angles of around 80 to 70$\degr$ have been inferred from observations \citep{lamy22}, corresponding to velocities of 0.17 to $0.34c$ (kinetic energies of 7.4 to 30 keV). From these values, the corresponding cone thickness ranges from $\sim10$ to $20\degr$. However, estimations from observations of the cone thickness imply values of around $1\degr$ \citep[e.g.][]{queinnec98, pachenko16}. It is currently unclear what the cause of this discrepancy is. With this in mind, we choose values for $\alpha$ and $\Delta\alpha$ independently of one another.

There may be certain configurations wherein the beams from both magnetic hemispheres are seen simultaneously. Assuming emission occurs in the same magnetoionic mode, the flux densities from each hemisphere will have opposite signs (neglecting the effects of radiative transfer). Therefore, in this scenario the circularly polarised flux density from each hemisphere will cancel one another out. However, the total flux density will still be received. In such a situation, we still consider the signal to be visible. Recall however that we limit emission to one hemisphere if the size of the field line exceeds $100~R_\star$ (see Section~\ref{sec:dipole}).

There are also a few conditions which must be satisfied for emission to be generated at the frequency $\nu$ at some point along the field line connecting the planet to the stellar surface. Firstly, the maximum cyclotron frequency on the field line $\nu_\textrm{max}$, which occurs at the footpoint of the field line, must be greater than $\nu$. Using Equation~\ref{eq:cyclotron frequency} and \ref{eq:field line strength - radius}, the maximum observable frequency is
\begin{equation}
\nu_\textrm{max} = 2.8 B_\star \Big(1 - \frac{3R_\star}{4L}\Big)^{1/2} ,
\label{eq:fmax}
\end{equation}
where $L$ is given by Equation~\ref{eq:field line size}. Similarly, the minimum frequency observable must exceed the minimum cyclotron frequency on the field line $\nu_\textrm{min}$, which occurs at the magnetic equator:
\begin{equation}
\nu_\textrm{min} = 1.4 B_\star \Big( \frac{R_\star}{L} \Big)^3 .
\end{equation}
The cyclotron frequency at the planet's position $\nu_\textrm{p}$ must also be considered:
\begin{equation}
\nu_\textrm{p} = 2.8 B_\star \Big(\frac{R_\star}{a} \Big)^3 \Big(1 - \frac{3a}{4L}\Big)^{1/2} .
\label{eq:fplanet}
\end{equation}
Provided $\nu_\textrm{max} > \nu > \nu_\textrm{min}$, and $\nu > \nu_\textrm{p}$, emission can occur in the hemisphere the planet occupies, as well as the opposite hemisphere provided $L < L_\textrm{max}$. However, if $\nu_\textrm{p} > \nu$, emission can only occur in the opposite hemisphere (again provided $L < L_\textrm{max}$), since no point on the star-planet field line in the hemisphere the planet occupies has a cyclotron frequency corresponding to the observing frequency.

\section{What orbital phases does emission appear at?}
\label{sec:orbital phases}

With the model described, we now demonstrate its applicability by illustrating the phenomenon of emission appearing at the quadrature points of a planet or satellite's orbit. These points correspond to orbital phases of around 0.25 and 0.75 (with 0 being primary conjunction). The phenomenon of enhanced radio emission from Jupiter at the quadrature points of Io's orbit was first identified almost six decades ago by \citet{bigg64}. As the magnetic SPI scenario represents an effectively scaled-up version of the Jupiter-Io system, there has been recent emphasis in the literature on detecting signatures of such interactions at radio wavelengths at quadrature points \citep[e.g.][]{perez-torres21, kavanagh22}.


\subsection{The expectation of emission at quadrature}
\label{sec:quadrature}

To understand the phenomenon of emission occurring near points of quadrature, consider the scenario where the orbital, rotation, and magnetic axes are all aligned along $\hat{z}$ in the plane of the sky (refer to Appendix~\ref{sec:star vectors} for definitions). In this `aligned' configuration, the planet orbits in the equatorial plane of the star, and its position as a function of orbital phase is described by the following vector (Appendix~\ref{sec:planet vectors}):
\begin{equation}
\hat{x}_\textrm{p} = \cos\phi_\textrm{p}\hat{x} + \sin\phi_\textrm{p}\hat{y} .
\end{equation}
The planet induces radio emission at the observing frequency along the field line connecting it to the star in both hemispheres. These points are $(r,\theta)$ and $(r,\pi - \theta)$ in the Northern and Southern hemispheres respectively. As the field line is symmetric about the magnetic equator, and the orbital distance is constant, the co-latitudes $\theta$ and $\pi - \theta$ always correspond to this frequency. The emission in each hemisphere is beamed in a cone centered along the vector
\begin{equation}
\hat{c} = \pm\Big(\frac{B_r}{B}\hat{r} + \frac{B_\theta}{B}\hat{\theta} \Big),
\end{equation}
where $\pm$ denotes the Northern/Southern hemisphere respectively, and $B_r$, $\hat{r}$, $B_\theta$, and $\hat{\theta}$ are defined in Equations~\ref{eq:Br} to \ref{eq:theta hat} (replacing $\hat{x}_\textrm{B}$ with $\hat{x}_\textrm{p}$ and $\hat{z}_\textrm{B}$ with $\hat{z}$ in this scenario). The direction of $\hat{c}$ relative to the line of sight determines if and when the emission is beamed towards the observer (Equation~\ref{eq:beam angle}). For emission from the Northern/Southern hemisphere to be seen twice per orbit, the angle between $\hat{c}$ and $\hat{x}$ ($\gamma$) must be within the range $\alpha\pm\Delta\alpha/2$ twice per orbit. Using Equations~\ref{eq:Br} to \ref{eq:theta hat}, in both magnetic hemispheres this angle can be shown to be
\begin{equation}
\cos\gamma = \frac{3\sin\theta\cos\theta}{(1 + 3 \cos^2\theta)^{1/2}}\cos\phi_\textrm{p} .
\label{eq:quadrature angle}
\end{equation}

\begin{figure}
\centering
\includegraphics[width = \columnwidth]{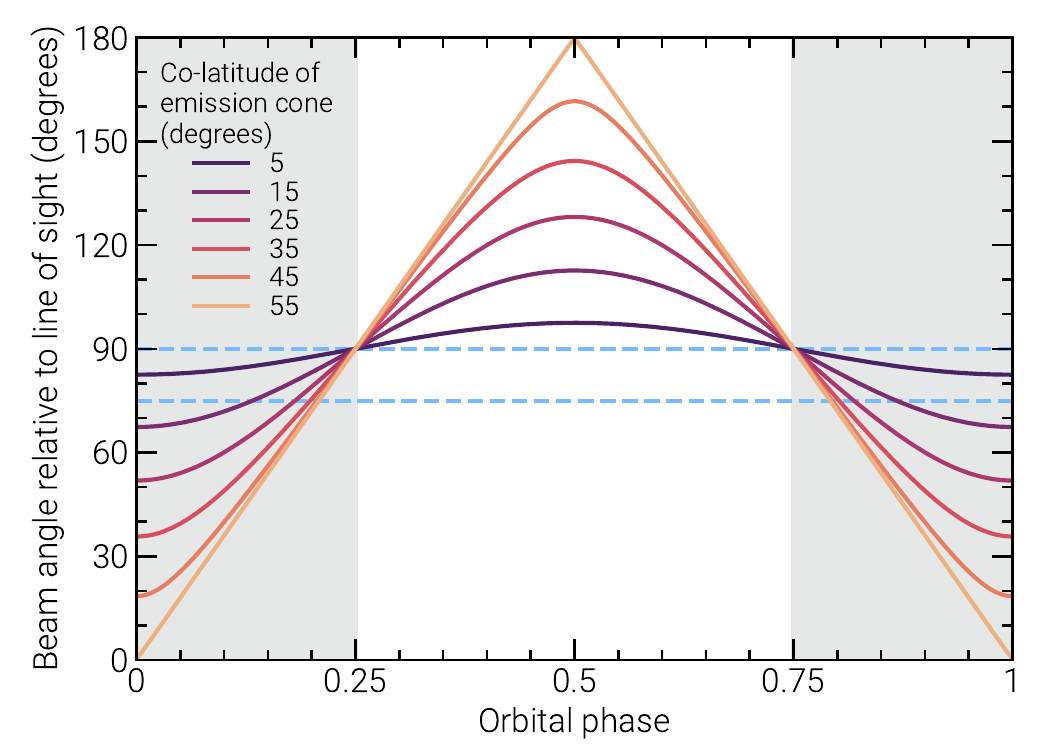}
\caption{The angle between the line of sight and the emission cone in the Northern magnetic hemisphere computed using Equation~\ref{eq:quadrature angle} at different magnetic co-latitudes ($\theta$) for the emitting point. When this angle equals the cone opening angle, emission is visible to the observer. Two blue dashed lines mark where the beam angle is $75\degr$ and $90\degr$ as examples. The system is oriented such that the rotation and magnetic axes of the star, as well as the orbital axis of the planet, are all aligned and lie in the plane of the sky. For a fixed co-latitude (emission frequency), the beam angle varies sinusoidally with the planet's orbital phase. The amplitude of the curve is maximised when $\theta = \cos^{-1}(1/\sqrt{3}) \approx 55\degr$. For cone opening angles less than $90\degr$, emission appears furthest from conjunction when $\theta=55\degr$. This can be seen in the example of when the beam angle equals $75\degr$, in that the orbital phases where it is visible are as far as possible from conjunction (an orbital phase of 0 or 1). At most, emission can be seen furthest from conjunction at orbital phases of 0.25 and 0.75, which are known as the quadrature points of the orbit. This however requires a cone opening angle of $90\degr$, which is the physical limit for the underlying emission mechanism (see text). Therefore, the shaded regions mark the range of orbital phases where emission can be considered to be in quadrature, assuming an aligned geometry for the system. Note that in aligned configurations, the results are equivalent for the Southern magnetic hemisphere, being invariant for the co-latitudes $\theta$ and $\pi - \theta$.}
\label{fig:quadrature angles}
\end{figure}

In aligned scenarios, Equation~\ref{eq:quadrature angle} tells us that emission at a given frequency is visible from both hemispheres simultaneously, assuming fixed parameters for the emission cone. What this frequency is depends on the values of $B_\star$ and $a$. In Figure~\ref{fig:quadrature angles}, we show $\gamma$ versus the orbital phase of the planet for different magnetic co-latitudes of the emitting point in the Northern hemisphere. We see that $\gamma$ varies between a minimum and maximum at primary ($\phi_\textrm{p} = 0$) and secondary transits ($\phi_\textrm{p} = 0.5$), with the amplitude of these curves being determined by the quantity $3\sin\theta\cos\theta / (1 + 3\cos^2\theta)^{1/2}$. The larger this quantity is, the further from primary transit the angle $\gamma$ is within the range $\alpha\pm\Delta\alpha/2$. It is maximised when $\theta = \cos^{-1}(1 / \sqrt{3}) \approx 55\degr$, giving $\cos\gamma = \cos\theta_\textrm{p}$. Therefore, the furthest emission can appear from primary transit is centered at orbital phases of $\alpha$ and $1 - \alpha$, which for the maximum value of $\alpha$ being $90\degr$, correspond to orbital phases of 0.25 and 0.75. 

In general, planet-induced radio emission in aligned systems is visible at orbital phases of $0$ to $(\alpha + \Delta\alpha / 2)$ and $1-(\alpha + \Delta\alpha / 2)$ to $1$. These phase intervals therefore set the minimum and maximum phases where emission can be considered to be at quadrature, with the exact phases being determined by the magnetic co-latitude of the emitting point. The minimum value of $\gamma$ is $\cos^{-1}(3\sin\theta\cos\theta / (1 + 3\cos^2\theta)^{1/2})$, which occurs at primary transit. To appear at least once, this quantity must be in the range $\alpha\pm\Delta\alpha/2$, and for emission to appear twice per orbit, it must be less than $\alpha - \Delta\alpha/2$.

\subsection{Signal visibility for aligned and misaligned systems}
\label{sec:aligned misaligned}

As shown in the previous section, planet-induced radio emission is always visible twice per orbit in aligned systems near the quadrature points of the planet's orbit, provided that the emission occurs at a magnetic co-latitude that is not close to $0\degr$ or $90\degr$. But what happens when the magnetic, rotation, and orbital axes are no longer aligned? 

\begin{figure}
\centering
\includegraphics[width = \columnwidth]{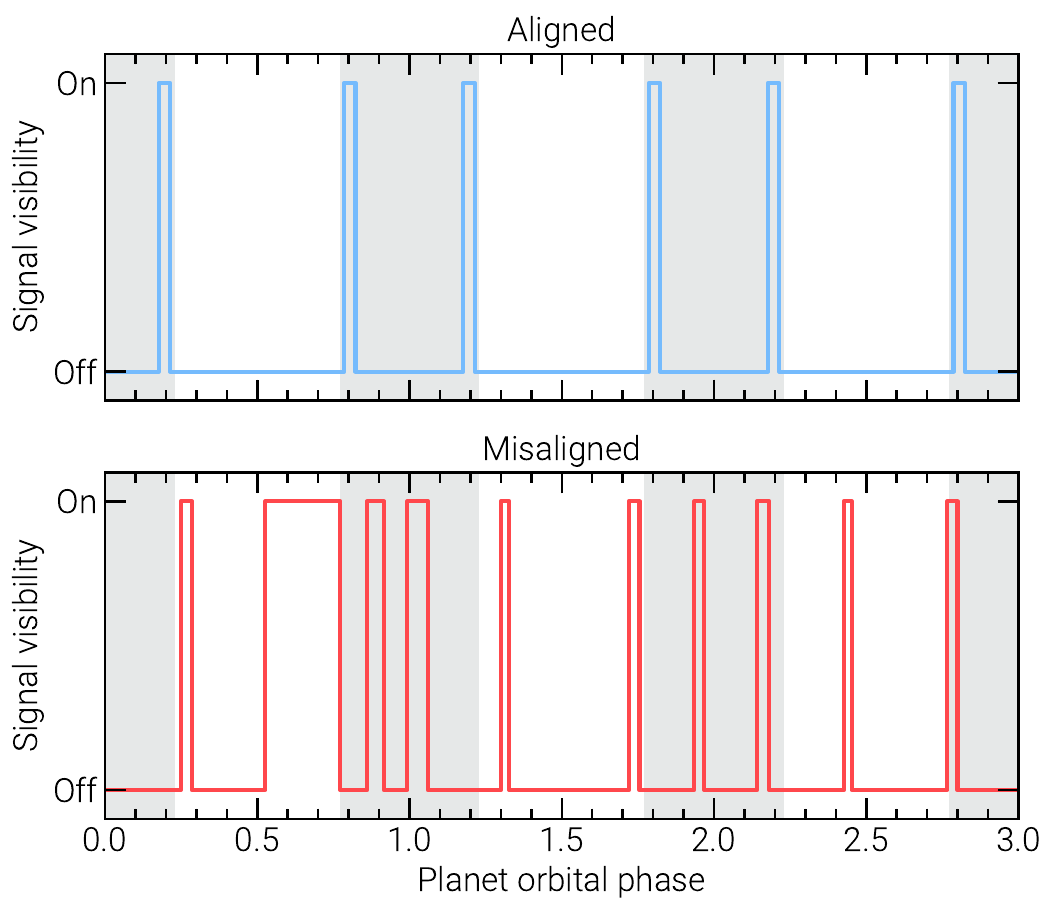}
\caption{Comparison of the visibility of planet-induced radio emission in an aligned (top) and misaligned (bottom) exoplanetary system for the first three orbits of the planet. The specific parameters in each case are listed in Table~\ref{tab:quadrature values}. The grey shaded regions in both panels illustrates the range of orbital phases where emission can be considered to be in quadrature (see text). In the misaligned scenario, emission regularly appears outside of quadrature.}
\label{fig:quadrature}
\end{figure}

In Figure~\ref{fig:quadrature}, we show what we refer to as `visibility lightcurves' for two configurations: an aligned (described in Section~\ref{sec:quadrature}) and a `misaligned' case.For the parameters in each case, see Table~\ref{tab:quadrature values}. These curves show a signal which is either `on' or `off'. When $\gamma$ is in the range $\alpha\pm\Delta\alpha/2$ in either hemisphere, we say the signal is `on', i.e. the emission can in theory be seen by the observer. Otherwise, the signal is `off'. This means that either the emission is not beamed along the line of sight at that time, or emission at the observing frequency cannot be generated at that time (see Equations~\ref{eq:fmax} to \ref{eq:fplanet}).

As can be seen from Figure~\ref{fig:quadrature}, in the aligned case, emission appears at the same orbital phases every orbit, near the quadrature limits described in Section~\ref{sec:quadrature}. However, in the misaligned scenario, this is no longer the case. For the first orbit of the planet, emission appears three times, two of which being outside of the range of possible quadrature phases. In the second orbit, it is seen four times, twice outside of quadrature. Finally, for the third orbit emission appears three times, once outside of quadrature. This demonstrates how complex morphology arises in lightcurves when the system is no longer aligned, resulting in emission appearing outside of quadrature for a significant amount of time. Note also that in the misaligned case, the time duration of each `on' window varies significantly.

\begin{figure}
\centering
\includegraphics[width = \columnwidth]{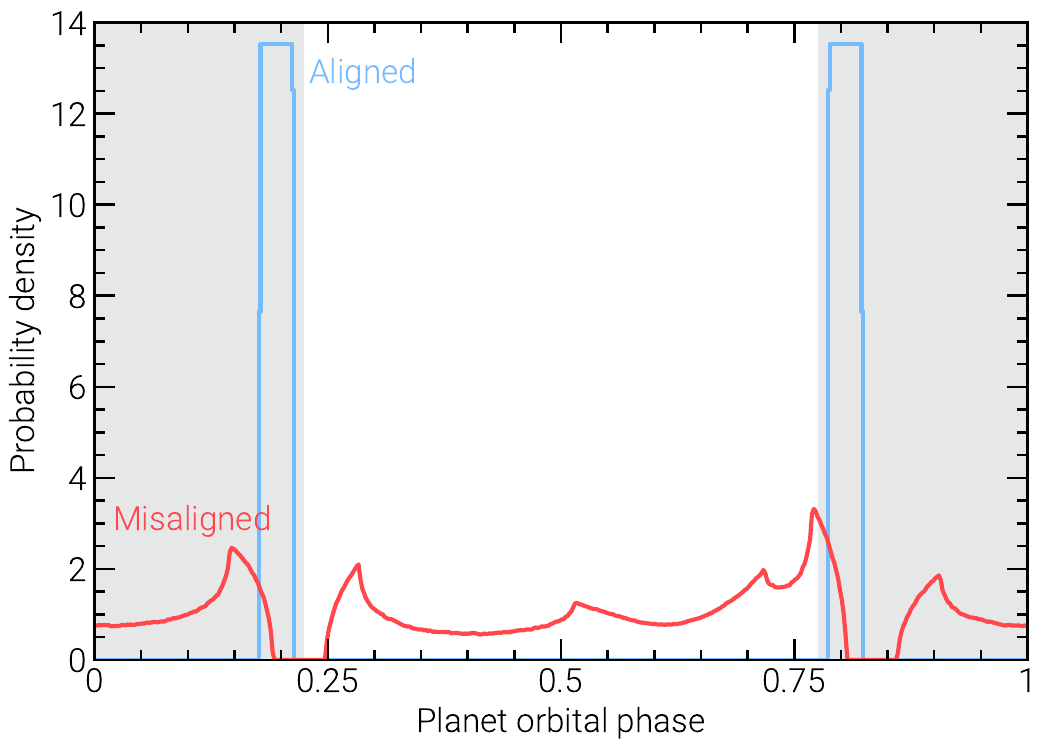}
\caption{Probability density of the orbital phases emission appears at in the aligned and misaligned scenarios. The integral of each curve over a given orbital phase interval gives the probability that visible emission occurs within that interval. The grey shaded regions again show the range of possible phases where emission can be considered to be in quadrature in the aligned scenario. In the misaligned scenario, the majority of the emission appears outside of quadrature.}
\label{fig:quadrature pdf}
\end{figure}

To further illustrate the significant differences between the emission morphology for aligned and misaligned systems, we compute the visibility lightcurves for each scenario for 500 orbits of the planet with 500 time elements per orbit, using the same parameters listed in Table~\ref{tab:quadrature values}. We then take the orbital phases where emission is visible and fold them with the orbital period of the planet, and compute the probability density (PD) of the visible emission as a function of orbital phase. These are shown in Figure~\ref{fig:quadrature pdf}. Unsurprisingly, in the aligned case emission is contained entirely within two narrow windows within the range of possible quadrature phases. In the misaligned case however, the distribution is much flatter, and has a significant component in the range of orbital phases associated with quadrature emission. Integrating the probability density in the misaligned case outside of quadrature (orbital phases of $\alpha+\Delta\alpha/2$ to $1 -\alpha-\Delta\alpha/2$), we find that emission appears outside of quadrature 57\% of the time. This illustrate that carrying out targeted radio observations of systems only at points of quadrature when we have little knowledge of the geometrical properties of the planetary orbit, magnetic field, and rotation axis of the star may not be the most appropriate course of action. Similarly, interpreting radio emission away from quadrature as being unrelated to SPI is also fraught. One must therefore use a geometric model such as the one presented in this work for analysis.

The fact that the Io induced emission on Jupiter appears almost exclusively at the quadrature points of Io's orbit \citep{marques17} is due to the fact it resembles the aligned configuration described here. This is because we view the system from the ecliptic plane of the solar system. We show this in Appendix~\ref{sec:io}, where we compare the probability density of emission in an aligned configuration to the results reported by \citet{marques17}.

\begin{table}
\caption{System parameters for the aligned and misaligned exoplanetary systems presented in Figures~\ref{fig:quadrature} and \ref{fig:quadrature pdf}.}
\label{tab:quadrature values}
\centering
\begin{tabular}{ccc}
\hline
Parameter & \multicolumn{2}{c}{Value} \\
\hline
$M_\star$ & \multicolumn{2}{c}{0.2~$M_{\sun}$} \\
$R_\star$ & \multicolumn{2}{c}{0.3~$R_{\sun}$} \\
$P_\star$ & \multicolumn{2}{c}{0.8~days} \\
$\phi_{\star,0}$ & \multicolumn{2}{c}{0} \\
$B_\star$ & \multicolumn{2}{c}{1~kG} \\
$a$ & \multicolumn{2}{c}{10~$R_\star$} \\
$\phi_{\textrm{p},0}$ & \multicolumn{2}{c}{0} \\
$\nu$ & \multicolumn{2}{c}{100~MHz} \\
$\alpha$ & \multicolumn{2}{c}{75$\degr$} \\
$\Delta\alpha$ & \multicolumn{2}{c}{10$\degr$} \\
& \underline{Aligned} & \underline{Misaligned} \\
$i_\star$ & 90$\degr$ & 52$\degr$ \\
$\beta$ & 0$\degr$ & 21$\degr$ \\
$i_\textrm{p}$ & 90$\degr$ & 67$\degr$ \\
$\lambda$ & 0$\degr$ & 36$\degr$ \\
\hline
\end{tabular}
\end{table}

\subsection{A departure from emission at quadrature}

\begin{figure*}
\centering
\includegraphics[width = \textwidth]{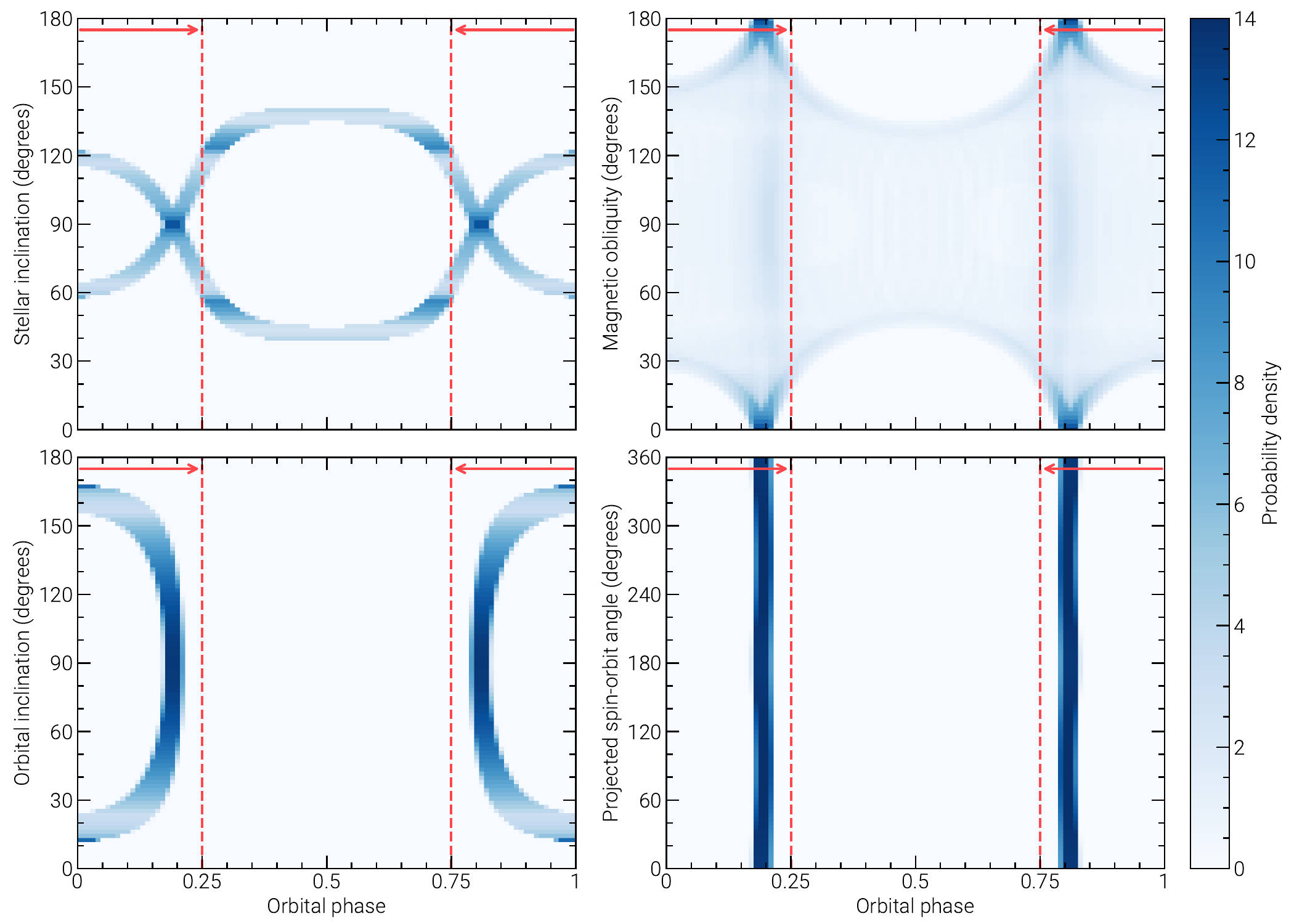}
\caption{Same as Figure~\ref{fig:quadrature pdf}, but varying each value of $i_\star$, $\beta$, $i_\textrm{p}$ and $\lambda$. In each Figure, the remaining values are those listed for the aligned case in Table~\ref{tab:quadrature values} (i.e. in the top left panel, $\beta = 0\degr$, $i_\textrm{p} = 90\degr$, and $\lambda = 0\degr$). Red arrows indicate the range of orbital phases where emission can be considered to be in quadrature, the limits of which are marked with vertical dashed lines.. The probability density is shown such that integrating horizontally over a given orbital phase interval gives the probability that observed emission occurs in that interval. Note that if the stellar or orbital inclination are low ($i_\star\lesssim20\degr$ or $\gtrsim160\degr$, $i_\textrm{p} \lesssim10\degr$ or $\gtrsim170\degr$), the emission is never visible for systems described by the remaining set of parameters.}
\label{fig:pdf angles}
\end{figure*}

We now explore the effects of each of the angles $i_\star$, $\beta$, $i_\textrm{p}$ and $\lambda$ on the PD of the lightcurve, to determine their effects on the range of orbital phases that emission can apppear at. Again we fix the remaining values as listed in Table~\ref{tab:quadrature values}, and vary $i_\star$, $\beta$, $i_\textrm{p}$ and $\lambda$ individually. Our resolution for $i_\star$, $\beta$, and $i_\textrm{p}$ is $1.8\degr$, and $3.6\degr$ for $\lambda$. We compute the lightcurve in the same manner as in Section~\ref{sec:aligned misaligned}, and then compute the PD of the emission as a function of orbital phase. The results of this are shown in Figure~\ref{fig:pdf angles}.

We see that apart from the projected spin-orbit angle, when the values of $i_\star$, $\beta$, and $i_\textrm{p}$ depart from those describing an aligned configuration, emission no longer primarily appears near the points of quadrature. Therefore, without knowledge of these parameters, scheduling radio observations at the quadrature points of a planet's orbit can result in limited or no visibility of the emission induced by the planet. The converse is also true. If we know these properties, the model provided here can be used to estimate what orbital phases to sample. 

We see that if the stellar or orbital inclination is low ($i_\star\lesssim20\degr$ or $\gtrsim160\degr$, $i_\textrm{p} \lesssim10\degr$ or $\gtrsim170\degr$), emission is never visible regardless of the magnetic obliquity and projected spin-orbit angle for systems described by the remaining parameters listed in Table~\ref{tab:quadrature values}. \citet{hess11} found a similar result using the ExPRES code, in that effectively zero planet-induced emission is seen in the systems they simulated for orbital inclination $\lesssim30\degr$ or $\gtrsim150\degr$, irrespective of the magnetic obliquity. Note that their analysis was limited to orbital inclinations and magnetic obliquities in increments of $15\degr$.

\section{What exoplanets are we biased towards detecting in the radio?}
\label{sec:bias}

One of the main motivators for developing the model presented here is to determine if we are biased towards detecting planet-induced radio emission from exoplanetary systems with certain architectures. Analogous of our detection bias towards orbits with $i_\textrm{p} \sim 90\degr$ when using the radial velocity and transit methods, certain orbital configurations may result in the induced ECM emission being beamed towards the observer for a longer duration of time (higher duty cycle) compared to other configurations. If this is the case, then systems identified as candidates for magnetic SPI via blind radio surveys may be more likely to reflect such configurations \citep[e.g.][]{callingham21b}.

To answer this question, we need to compute the visibility lightcurves for a wide range of parameters, and determine which parameters (if any) produce emission with a high duty cycle. As there are a large number of parameters (Table~\ref{tab:model parameters}), we choose random samples for each one. Next, we describe our choices for the range of values for each parameter, as well as the underlying distribution we draw them from.


\subsection{The parameter space for planet-hosting M~dwarfs}
\label{sec:samples}

As M~dwarfs are likely to be the most favourable targets for detection of planet-induced radio emission, we focus on sampling a parameter space reflective of these stars. The masses of M~dwarfs range from $\sim0.1$ to 0.6~$M_{\sun}$, and volume-limited surveys of nearby M~dwarfs suggest that the number of M~dwarfs drops off linearly with mass, with late-type M~dwarfs being around four times as common as early-types \citep{winters19}. Therefore, we draw samples for the stellar mass from a linear distribution with the same slope as that found by \citet{winters19}. The mass and radii of M~dwarfs relate via \citep{schweitzer19}
\begin{equation}
R_\star = (0.935\pm0.015)M_\star + (0.0282\pm0.0068) ,
\label{eq:mass-radius relation}
\end{equation}
where $R_\star$ and $M_\star$ are in solar units. We use Equation~\ref{eq:mass-radius relation} to draw samples for the stellar radius based on the samples drawn for $M_\star$, assuming the errors in Equation~\ref{eq:mass-radius relation} are Gaussian. For masses of 0.1 to 0.6~$M_{\sun}$, the resulting radii range from $\sim0.1$ to 0.6~$R_{\sun}$. Note that this relation is derived from eclipsing binaries, which is assumed to hold for single stars \citep{schweitzer19}.

The rotation periods of M~dwarfs depends on both their spectral type (mass) and age \citep{popinchalk21, lu22}. For early M-stars, there is evidence for a bimodal distribution of rotation periods, which disappears past the fully-convective boundary. \citet{lu22} suggest that this is either due to these stars rapidly spinning down at around 3~Gyr, or a detection bias disfavouring stars with intermediate periods, which exhibit lower levels of variability and therefore are more difficult to measure rotation periods for \citep[see also][]{reinhold19}. In addition to these uncertainties, there are only a small number of late-M stars with measured rotation periods \citep{popinchalk21}. With this in mind, along with the fact that we do not explicitly consider the age/activity of the star, we opt to choose samples for the rotation period uniformly in the range of 0.1 to 160 days, which covers the rotation periods of the M~dwarfs presented by \citet{popinchalk21}. For the inclination of the rotation axis, there should be no preferential orientation of the vector $\hat{z}_\star$ when projected onto a unit sphere centered on the observer. Therefore, we  sample $\cos(i_\star)$ uniformly from 1 to $-1$ (0 to $180\degr$), which gives a uniform surface density of points on the unit sphere. For the initial rotation phase, we choose values from 0 to 1 uniformly.

The dipolar magnetic field strengths of M~dwarfs are estimated to range from at least 100~G to a few~kG depending on their activity. Such information along with the magnetic obliquity can be inferred with ZDI. M~dwarfs exhibit a range of surface field magnetic configurations. Going from early to mid-type M~dwarfs, their fields transition from being relatively weak and non-axisymmetric \citep[that is each magnetic multipole is not aligned with the rotation axis;][]{donati08} to being strong and axisymmetric \citep{morin08}, resembling aligned dipoles. Interestingly, late-M stars appear to exhibit both configurations \citep{morin10}. There is a further complication to this. ZDI generally only recovers a fraction of the underlying magnetic energy, which depends on the magnetic multipole. This fraction of energy recovered by ZDI also depends on both the inclination of the stellar rotation axis and the rotation rate of the star \citep[see][]{lehmann19, lehmann21}. Note however this has only been studied in the context of the Sun, as we cannot assess the true magnetic topology of other stars. What is clear however from Figure~12(c) of \citet{lehmann19} is that this effect is most severe for the dipolar component of the magnetic field. In short, ZDI can provide information about the strength and obliquity of the dipole component of the magnetic fields of M~dwarfs. However, depending on the spectral type, inclination, and rotation period, its true strength may be difficult to recover with ZDI. With this in mind, as well as the fact that the dipole field strengths and obliquities are not generally explicitly stated in the literature, we again take an uninformed approach and draw the samples for the dipole field strength and obliquity from uniform distributions. For the field strengths, we consider values from 100~G to 1~kG, and for the obliquity, 0 to $180\degr$.

In terms of the planet itself, we can first impose a lower limit for its orbital distance using the Roche limit, which tells us the minimum distance a planet can be to its host star before it starts to disintegrate. Massive, gaseous exoplanets are more susceptible to this compared to rocky exoplanets. Therefore, the shortest period planets around stars are likely to be rocky. For incompressible bodies (i.e. rocky planets), the Roche limit for its orbital distance is \citep{rappaport13}
\begin{equation}
\frac{a}{R_\star} > 2.44\Big( \frac{\rho_\star}{\rho_\textrm{p}} \Big)^{1/3} ,
\end{equation}
where $\rho_\star$ and $\rho_\textrm{p}$ are the densities of the star and planet. The density of the star is $\rho_\star = 3 M_\star / 4\pi {R_\star}^3$, and the lower limit for the orbital distance as a function of stellar mass is smallest when the planet density is highest. For rocky planets, this is estimated to be around 8~g~cm$^{-3}$ \citep{unterborn19}. Therefore
\begin{equation}
\frac{a}{R_\star} > 0.75 {M_\star}^{1/3} {R_\star}^{-1} .
\end{equation}
Note that $M_\star$ and $R_\star$ are in CGS units here.

The relevant outer limit for the orbital distance in the context of magnetic SPI on M~dwarfs is the size of the Alfv\'en surface. Outside this region, the planet cannot induce radio emission from the star. Therefore, we set the upper limit for the orbital distance as the maximum radius of the Alfv\'en surface. This generally corresponds to where the magnetic field lines begin to open, which in our model we set to occur at 100 stellar radii, so we adopt the same value for the upper limit. This value is consistent with MHD models of the wind of WX~UMa \citep{kavanagh22}, which possesses one of the strongest magnetic fields measured to date \citep{shulyak17}. \citet{kavanagh22} estimated the size of the Alfv\'en surface to be around 80 stellar radii. Note however that an Alfv\'en surface of this size is likely only valid for the most active M~dwarfs, and is likely an overestimation in the case of inactive M~dwarfs. However, this information cannot be determined without some form of stellar wind modelling.

With these limits in place, the next question is what distribution to choose for the orbital distances. In general, it is easier to find planets the closer they orbit to their host star. Additionally, larger planets are also more easily detected. On top of this, formation models are presently at odds with the observed exoplanet demographics for M~dwarfs. So far, more massive and fewer short-period (small orbital distance) planets have been found around M~dwarfs compared to what these models predict \citep[see][]{schlecker22, ribas23}. Given these uncertainties, we again opt for a uniform distribution for the orbital distance.

If the orbital axis $\hat{z}_\textrm{p}$ is independent of the rotation axis $\hat{z}_\star$, the distribution of orbital inclinations should be uniform in $\cos i_\textrm{p}$ such that the tips of the orbital axes are uniformly distributed over a unit sphere. This distribution combined with a uniform distribution for $\cos i_\star$ results in distribution for the true spin-orbit angle $\psi$ that is uniform in $\cos\psi$. The corresponding distribution for the projected spin-orbit angle $\lambda$ is also uniform. Observations hint at an underlying bimodal distribution of spin-orbit angles centered at $\psi\approx0\degr$ and $90\degr$ \citep{stefansson22, albrecht22}. If that is the case, then clearly there must be some relationship between the direction of $\hat{z}_\star$ and $\hat{z}_\textrm{p}$. However, the number of measurements for $\psi$ and $\lambda$ are limited, particularly for M~dwarfs, and can only be measured for transiting exoplanets. Due to these low numbers, we opt for an uninformed approach, and uniformly sample values for $\cos i_\textrm{p}$ from -1 to 1 and for $\lambda$ from 0 to $360\degr$. For the initial orbital phase of the planet, we also uniformly sample the values from 0 to 1.

The final set of values to sample relate to the emission. For fundamental cyclotron emission, the upper limit for the observing frequency is set by the maximum field strength we consider, which is 1 kG. The corresponding cyclotron frequency for this field strength is 2.8~GHz via Equation~\ref{eq:cyclotron frequency}. For the lower limit of the observing frequency, we set this to 10~MHz, which is the lowest operating frequency of current-generation radio telescopes \citep[e.g. LOFAR,][]{edler21}. Again, it is not clear what the underlying distribution of emitted frequencies is, given that there has yet to be a conclusive detection of such emission. Furthermore, a sophistication model for the evolution of the velocity distribution of the electrons powering the maser as they travel along the field line is required to accurately determine the frequencies at which the emission occurs over time. Lacking this information, we once again uniformly sample the observing frequency between 10~MHz to 2.8~GHz.

For the properties of the emission cone, we adopt a range of 70 to $80\degr$ for the opening angle based on the discussion point in Section~\ref{sec:radio}. Similarly for the cone thickness, we are limited to the Jupiter-Io interaction in terms of our knowledge of appropriate values. While observations suggest thicknesses of around $1\degr$, theoretical considerations suggest values of 10 to $20\degr$ based on the range of observed opening angles. To not overestimate the thickness more than necessary, we set the upper limit to $10\degr$, and the lower limit to $1\degr$. Again, the lack of observations and a sophisticated model for the maser limit our ability to implement meaningful ranges and distributions for the cone properties in a stellar context. Our focus here however is to evaluate the geometric dependence of the duty cycle. As such, we employ uniform distributions for both values. Future work that better-establishes what are appropriate values and distributions for these quantities will allow for this to be re-assessed.


\subsection{Temporal resolution of the lightcurve}

An important aspect to consider here is the temporal resolution $\Delta t$ of the visibility lightcurve. Generally, for systems with short orbital periods (small orbital distances) and narrow cone thicknesses are only visible for very short windows. If $\Delta t$ is too large, we can end up undersampling and missing a large fraction of the on phases of the signal. We can determine suitable values for $\Delta t$ however by considering the time it takes for the emission cone to sweep across the line of sight. We approximate this as the duration of time taken for the planet to increase in orbital phase by $\Delta\alpha$ (the cone thickness), which is $P_\textrm{p}(\Delta\alpha / 360\degr)$. With the aim of resolving each on window with at least two points, we compare the duty cycle for a few hundred random samples for a signal duration of 1000~days, using time intervals of $\Delta t = P_\textrm{p}(\Delta\alpha / 720\degr)$ and $\Delta t = P_\textrm{p}(\Delta\alpha / 36000\degr)$. We find that the duty cycle obtained using the lower resolution varies by less than 4\% compared to the high resolution calculation. We therefore determine that $\Delta t = P_\textrm{p}(\Delta\alpha / 720\degr)$ is a suitable resolution to dynamically set for each lightcurve such that the true duty cycle of the signal is recovered.


\subsection{What systems are easiest to detect?}
\label{sec:bias general}

With the assumed parameter space of planet-hosting M~dwarfs laid out, we now perform a Monte Carlo simulation, sampling each parameter from their aforementioned distributions (Section~\ref{sec:samples}). We choose 1 million values for each parameter. For each set of values, we first compute their visibility lightcurves, and then their duty cycle (the percentage of time the signal is visible for). The time duration of each lightcurve is 500 days. We find that a randomly sampled system has on average a duty cycle of 4\%, and that 48\% of all systems can produce emission that is ever visible. In other words, 52\% of systems will never be observable, assuming static conditions for the large-scale magnetic field of the star and the emission cone. Of the 48\% of systems visible, their average duty cycle is 8\%. We also find that emission is as likely to be seen from the Northern magnetic hemisphere as the Southern magnetic hemisphere. This is unsurprising, as unless there is some special configuration of the system, the planet will spend as much time in the Northern hemisphere as the Southern hemisphere. In other words, there is no preferential polarisation for the radio emission, assuming both hemispheres emit via the same magnetoionic mode.

We then investigate each of the parameters to see which (if any) enhance the duty cycle, and if so, what values of the parameters do so. Due to both the high number of dimensions of the model and the random sampling, there is a large amount of scatter when plotting the duty cycle against each parameter for all of our samples. Each scatter plot is shown in Appendix~\ref{sec:scatter plots}. However, we see that there are certain values for the stellar inclination, magnetic obliquity, and orbital inclination which result in high duty cycles.


\subsubsection{Stellar parameters}
\label{sec:optimal - star}

\begin{figure}
\centering
\includegraphics[width = \columnwidth]{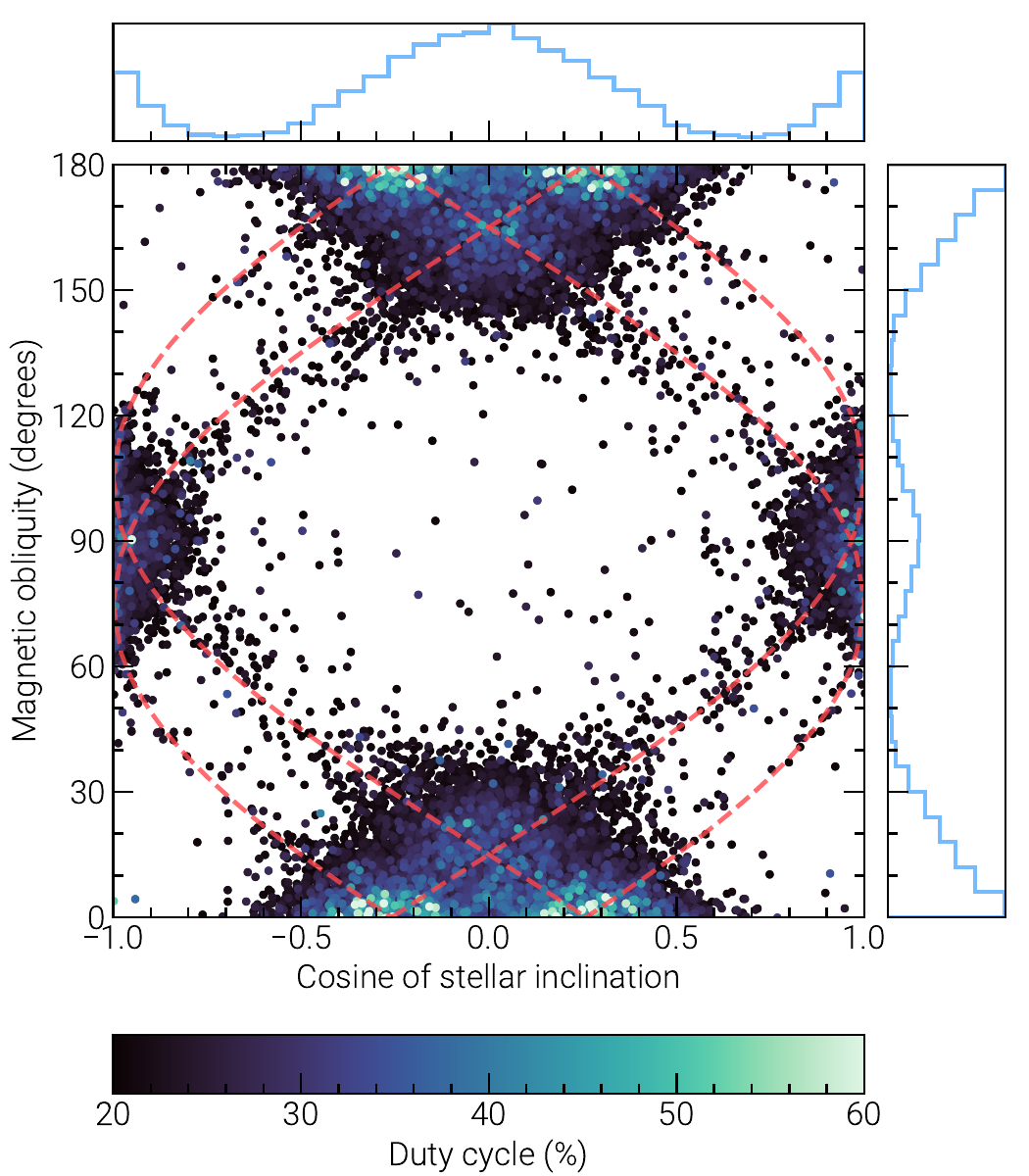}
\caption{Scatter plot of the magnetic obliquity ($\beta$) against the cosine of the stellar inclination ($i_\star$) for the systems where the duty cycle is greater than 20\%. Each point is coloured with its respective duty cycle, which are plotted in order of increasing duty cycle. We also show two sub-panels on the top and right, which are the normalised histograms of the points along each axis. Each histogram has 30 bins. The dashed red lines show where $\beta \pm i_\star$ is either $\alpha$ or $180\degr - \alpha$, where $\alpha$ is the opening angle of the emission cone. For clarity, we show lines for $\alpha = 75\degr$, which is the middle value of those we consider. These configurations result in the magnetic axis being able to form the angle $\alpha$ with the line of sight. The most visible systems are those where the magnetic axis is always tilted relative to the line of sight by $\alpha$. There are two distinct sets of values that result in this configuration, which we refer to as C1 and C2 (see text).}
\label{fig:scatter star}
\end{figure}

We first consider the stellar parameters that produce high duty cycles. In Figure~\ref{fig:scatter star}, we plot the magnetic obliquity against the cosine of the stellar inclination for the systems where the duty cycle exceeds 20\%. We see that the majority of the points all lie along curved lines, and the most visible systems are described by two distinct configurations. We refer to these as C1 and C2. In C1, the rotation axis forms the angle $\alpha$, the cone opening angle, with the line of sight ($i_\star = \alpha$ or $180\degr - \alpha$), and the magnetic axis is either parallel or anti-parallel to the rotation axis ($\beta = 0\degr$ or $180\degr$). In C2, the rotation axis is viewed pole on ($i_\star = 0\degr$ or $180\degr$), and the magnetic obliquity is $\alpha$ or $180\degr - \alpha$.

We can understand the structure seen in Figure~\ref{fig:scatter star}, as well as the high duty cycles of C1 and C2 by considering the angle $\chi$ that the magnetic axis $\hat{z}_\textrm{B}$ makes with the line of sight $\hat{x}$. Using Equations~\ref{eq:z_s} to \ref{eq:z_B}, we can express $\chi$ as:
\begin{equation}
\cos \chi = \hat{z}_\textrm{B} \cdot \hat{x} = \cos i_\star \cos \beta + \sin i_\star \sin \beta \cos \phi_\star .
\label{eq:magnetic axis inclination}
\end{equation}
The values for $\cos\chi$ are always in the range $\cos(i_\star \pm \beta)$. Overlaying different values of $i_\star \pm \beta$ onto Figure~\ref{fig:scatter star}, we find that the vast majority of the systems follow the lines where $i_\star \pm \beta$ is either $\alpha$ or $180\degr - \alpha$. In other words, systems with high duty cycles are those where the magnetic axis can form the angle $\alpha$ with the line of sight.

Equation~\ref{eq:magnetic axis inclination} also explains the most visible systems described by C1 and C2. If $\sin i_\star$ or $\sin\beta$ are zero, the angle $\chi$ no longer has any time dependence, and the magnetic axis is \textit{always} inclined relative to the line of sight by the same angle. The key distinction between C1 and C2 is that in C1, the magnetic axis remains fixed in place from an observer's point of view. In C2 however, the magnetic axis precesses about the line of sight. 90\% of all systems with duty cycles exceeding 40\% are in C1, and 3\% are in C2 (with a tolerance of $\pm10\degr$ for the inclination and obliquity), and the max duty cycles in C1 and C2 are 80 and 60\% respectively. The lower number of systems in C2 is primarily due to adopting a uniform distribution for the stellar inclination axes (Section~\ref{sec:samples}), which results in pole-on systems being much rarer than the near-equator on systems described by C1.

\subsubsection{Planetary parameters}
\label{sec:optimal - planet}

\begin{figure}
\centering
\includegraphics[width = \columnwidth]{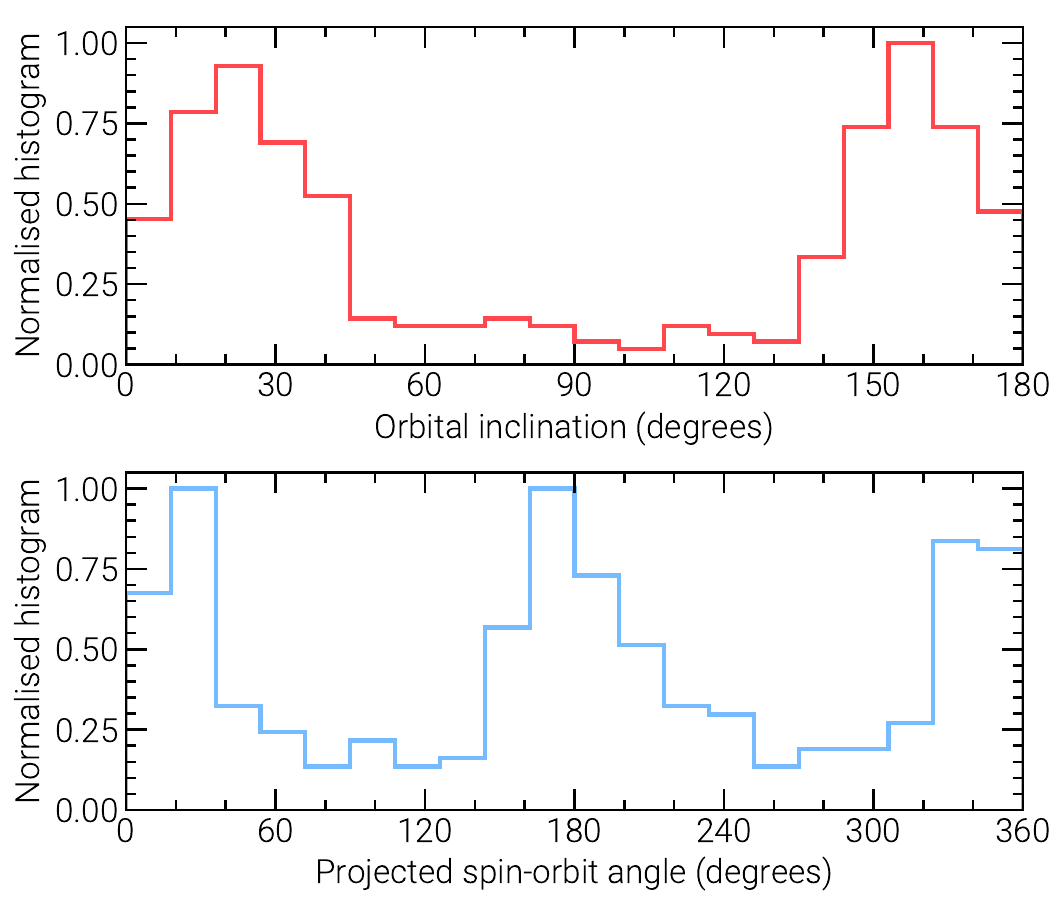}
\caption{Normalised histograms of the number of systems where the duty cycle of induced radio emission exceeds 40\% as a function of the orbital inclination (top) and projected spin-orbit angle (bottom). Both histograms have 20 equal width bins.}
\label{fig:planet hists}
\end{figure}

\begin{figure*}
\centering
\includegraphics[width = \textwidth]{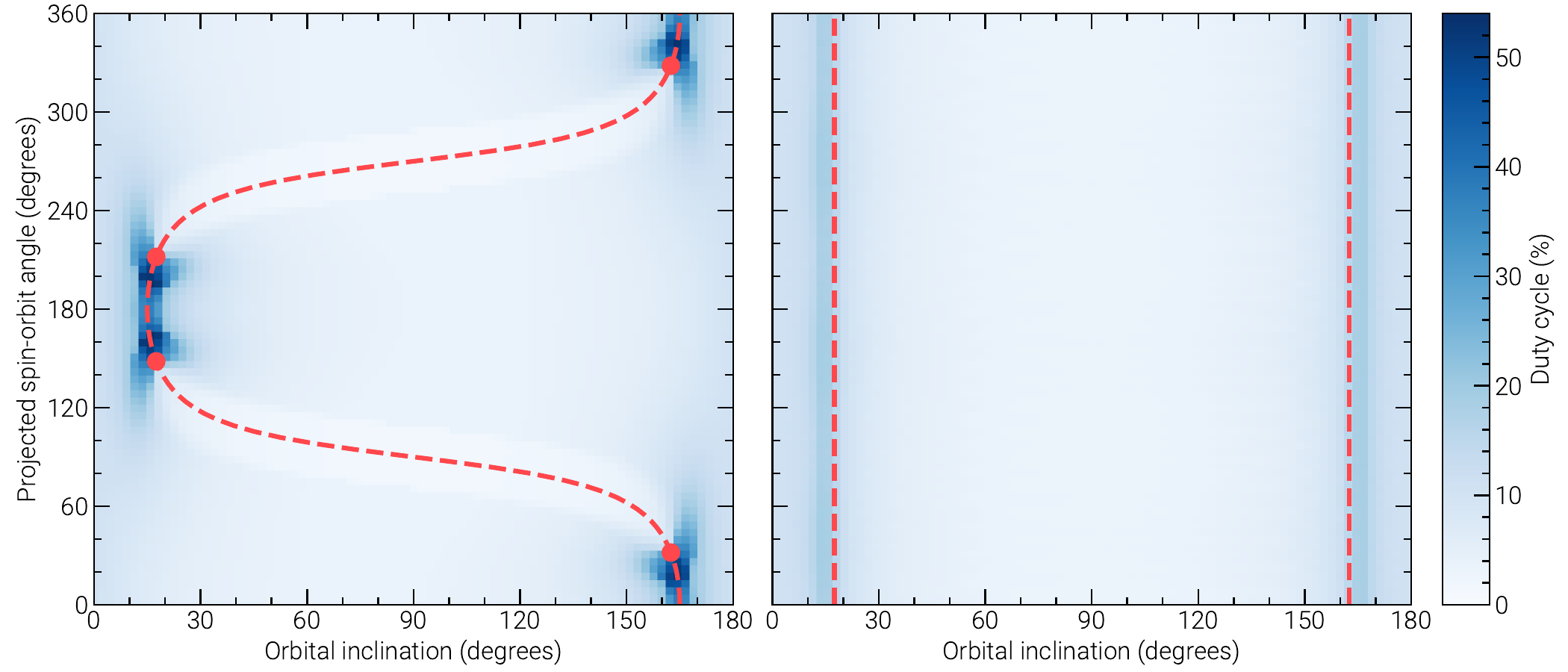}
\caption{The duty cycle of emission at 100~MHz from the Northern magnetic hemisphere of a star in C1 ($i_\star = \alpha = 75\degr$, $\beta = 0\degr$, left panel) and C2 ($i_\star = 0\degr$, $\beta = \alpha = 75\degr$, right panel) as a function of the planet's orbital inclination and projected spin-orbit angle. The cone thickness is $5\degr$, and the remaining parameters are those listed in Table~\ref{tab:quadrature values}. For C1, we see that the maximum duty cycle corresponds to orbits that pass over the magnetic poles, which are indicated by the dashed line. Note that only certain orbits that pass over the magnetic poles result in high duty cycles, which are marked with red circles (see text). For C2, the magnetic axis cannot always stay aligned with the orbital plane. As a result, the duty cycle has no dependence on the projected spin-orbit angle. It still peaks however at the same orbital inclinations as in C1, which are shown with vertical dashed lines.}
\label{fig:c1-c2}
\end{figure*}

We now must also consider the orbit of the planet around the star when it is in C1 or C2 to understand the high duty cycles seen in Figure~\ref{fig:scatter star}. In Figure~\ref{fig:planet hists} we show normalised histograms of the number of systems with duty cycles exceeding 40\% as a function of orbital inclination and projected spin-orbit angle. We see that most systems are near face-on ($i_\textrm{p} \approx 0\degr$ or $180\degr$) and have projected spin-orbit angles of either $\approx0\degr$ or $180\degr$. To further explore this, in Figure~\ref{fig:c1-c2} we plot the duty cycle of emission induced at 100~MHz as a function of the planet's orbital inclination and projected spin-orbit angle, from the Northern hemisphere of a star in C1 ($i_\star = \alpha = 75\degr$, $\beta = 0\degr$) and C2 ($\beta = \alpha = 75\degr$, $i_\star = 0\degr$). We compute each lightcurve for 100 orbits, with 100 time samples per orbit. For C1, we find that the maximum duty cycles correspond to planetary orbits that pass over the magnetic poles. For this to occur, the normal to the orbital plane $\hat{z}_\textrm{p}$ must be perpendicular to the magnetic axis $\hat{z}_\textrm{B}$. In C1 ($i_\star = \alpha = 75\degr$, $\beta = 0\degr$), this requires that:
\begin{equation}
\hat{z}_\textrm{B} \cdot \hat{z}_\textrm{p} = \cos i_\textrm{p} \cos \alpha + \sin i_\textrm{p} \sin \alpha \cos \lambda = 0,
\end{equation}
meaning that
\begin{equation}
\tan i_\textrm{p} \cos\lambda = \frac{-1}{\tan\alpha}.
\label{eq:polar orbits}
\end{equation}

The dashed line in the left panel of Figure~\ref{fig:c1-c2} shows the combined values of $i_\textrm{p}$ and $\lambda$ that describe orbits which pass over the magnetic poles, satisfying Equation~\ref{eq:polar orbits}. This line intersects with the regions where the duty cycle peaks, which occur at $i_\textrm{p} \approx 15\degr$ and $\lambda \approx 160\degr$ or $200\degr$, and $i_\textrm{p} \approx 165\degr$ and $\lambda \approx 20\degr$ or $340\degr$. Not all orbits described by the dashed line have high duty cycles however, which implies that further constrains exist which likely relate to the fraction of the orbit where the emission cones point along the line of sight. This is not trivial to show analytically with an exact treatment of the geometry. However, since the planet orbits over the magnetic poles, the field lines it interacts with are almost entirely radial for a significant part of its orbit. This means that the emission cone vector $\hat{c}$ is parallel to the position vector of the planet $\hat{x}_\textrm{p}$. 

If we assume that the field lines are radial, the angle between the cone and line of sight in the Northern magnetic hemisphere is (Equation~\ref{eq:beam angle}):
\begin{equation}
\cos\gamma = \hat{x}_\textrm{p} \cdot \hat{x} = \sin i_\textrm{p} \cos \phi_\textrm{p} .
\label{eq:beam angle radial field lines}
\end{equation}
Near $i_\textrm{p} = 15\degr$ and $165\degr$, $\gamma$ varies sinusoidally, the minimum of $\gamma$ is close to $\alpha$ which occurs at conjunction ($\phi_\textrm{p} = 0$). If the minimum of $\gamma$ is $\alpha - \Delta\alpha/2$, then the range of orbital phases where $\gamma$ is within $\alpha\pm\Delta\alpha/2$ is maximised, resulting in the highest duty cycle possible. In other words, the duty cycle is maximised when
\begin{equation}
\cos(\alpha - \Delta\alpha/2) = \sin i_\textrm{p} ,
\end{equation}
i.e.~when $i_\textrm{p} = 90\degr - \alpha + \Delta\alpha/2$ or $90\degr + \alpha - \Delta\alpha/2$. For $\alpha = 75\degr$, we have $i_\textrm{p} = 17.5\degr$ and $162.5\degr$. There are two corresponding values of $\lambda$ for each of these orbital inclinations that describe orbits which pass over the magnetic poles, which are obtained from Equation~\ref{eq:polar orbits}. For $i_\textrm{p} = 17.5\degr$ we have $\lambda \sim 148.2\degr$ and $211.8\degr$, and for $i_\textrm{p} = 162.5\degr$ we have $\lambda \sim 31.8\degr$ and $328.2\degr$. These four orbital configurations closely align with the regions where the duty cycle peaks seen in Figure~\ref{fig:c1-c2}, which are indicated by red circles.

In C2, the magnetic axis cannot stay in the orbital plane due to its precession about the rotation axis of the star. That being said, the magnetic axis and orbital plane will become aligned twice per stellar rotation. For an example configuration of C2 ($i_\star = 0\degr$, $\beta = \alpha = 75\degr$), the magnetic axis and orbit normal are perpendicular when
\begin{equation}
\hat{z}_\textrm{B} \cdot \hat{z}_\textrm{p} = \cos i_\textrm{p}\cos\alpha - \sin i_\textrm{p}\sin\alpha\cos(\phi_\star+\lambda) = 0,
\end{equation}
i.e.~when
\begin{equation}
\phi_\star = \cos^{-1}\Big[\frac{1}{\tan i_\textrm{p}\tan\alpha}\Big] - \lambda~\textrm{or}~2\pi - \cos^{-1}\Big[\frac{1}{\tan i_\textrm{p}\tan\alpha}\Big] - \lambda .
\end{equation}
So, while the rotation phases where $\hat{z}_\textrm{B}$ and $\hat{z}_\textrm{p}$ become perpendicular depend on the projected spin-orbit angle $\lambda$, the magnetic axis will always align with the orbital plane twice per orbit irrespective of the value of $\lambda$, provided that $1 / \tan\alpha < \tan i_\textrm{p} < - 1 / \tan\alpha$. When $\hat{z}_\textrm{B}$ and $\hat{z}_\textrm{p}$ are perpendicular, Equation~\ref{eq:beam angle radial field lines} is then valid under the assumption that the field lines are radial. Following the same logic as for C1, the duty cycle is maximised when $i_\textrm{p} = 17.5\degr$ or $162.5\degr$. This is what is seen in the right panel of Figure~\ref{fig:c1-c2}.

An interesting result from these configurations is that they produce emission that is predominantly either right (RCP) or left circularly polarised (LCP). In other words, their emission comes from either the Northern or Southern magnetic hemisphere, irrespective of if the star is in C1 or C2. In fact, when the duty cycle exceeds 40\%, virtually all systems emit either RCP or LCP exclusively. This is because these two configurations require one magnetic pole to always face towards the observer. This feature has been identified in the dynamic spectra of radio bursts from a sample of M~dwarfs by \citet{villadsen19}, which is expected if the electrons powering the radio emission are accelerated in the large-scale magnetic field of the star as we model in this work.

We also note that we see a marginal bias towards detecting emission induced by closer in planets (see Figure~\ref{fig:scatter plots}). This is due to our assumption that the field lines become open when the planet interacts with a field line that connects near to the magnetic poles. If the planet orbits far from the star, virtually all the field lines it sees will be open, and as a result the visibility of the emission is only possible from a single magnetic hemisphere, marginally reducing the likelihood of seeing the emission.

\begin{figure}
\centering
\includegraphics[width = \columnwidth]{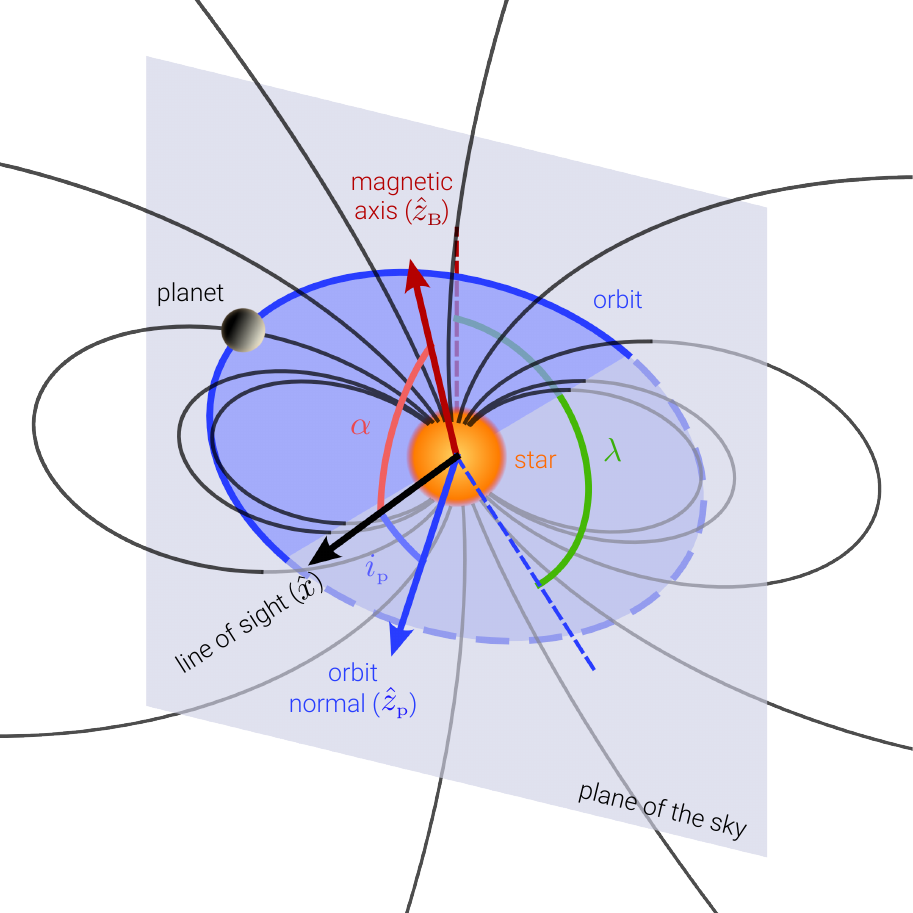}
\caption{Sketch of a planet orbiting around a star in the configuration C1 in the optimal configuration. The star has no magnetic obliquity here, and its magnetic axis is aligned with the stellar rotation axis, which itself is inclined relative to the line of sight by the angle $\alpha$, which is the angle radio emission induced on the star by the planet is beamed at from the magnetic field lines. The planet's orbit is shown in blue, and the black lines show the large-scale magnetic field of the star that connect to its orbit. The values of the orbital inclination $i_\textrm{p}$ and projected spin-orbit angle $\lambda$ are such that the duty cycle of radio emission induced by the planet on the star is maximised (see Section~\ref{sec:optimal - planet}). For the values of $\alpha$ considered in this work, the optimal orbital configuration is near face-on (i.e. the planet orbits in the plane of the sky).}
\label{fig:C1}
\end{figure}


\subsubsection{Emission parameters}

Aside from the geometrical parameters, the sampling also shows us that low-frequency emission from stars with strong magnetic fields are more favourable (Figure~\ref{fig:scatter plots}). This is unsurprising, as under the assumption of a uniform distribution of field strengths, lower frequencies are more likely as opposed to higher frequencies. Similarly, stars with stronger fields can produce a wider range of observable frequencies.

In terms of the cone properties, we see a marginal bias towards systems with cone opening angles closer to $90\degr$. For systems where the duty cycle exceeds 20\%, those which have opening angles close to $80\degr$ are about 1.5 times more likely to be seen than those where the opening angle is around $70\degr$. This can be understood by considering the configurations which we identify in Sections~\ref{sec:optimal - star} and \ref{sec:optimal - planet} that correspond to high duty cycles, which rely on the planet passing over the magnetic poles of the star. If the cone opening angle is closer to $90\degr$, then the emission cones are at right angles to the magnetic field. In a face-on orbit, this means that the cones always point towards the observer, assuming the magnetic axis lies in the orbital plane. We also see that thicker emission cones produce more visible emission. This is expected since a thicker cone results in wider windows wherein the signal can be seen.


\section{Detectability via other methods and prospects for transiting exoplanets}
\label{sec:transiting}

In the previous Section, we have identified two key configurations for the star and planetary orbit which result in planet-induced radio emission being visible for the majority of the time. That being said, these configurations describe planets in near face-on orbits, which are likely to be very difficult to detect via the radial velocity method, and also do not transit. These planets could theoretically be directly imaged if they orbit sufficiently far from their host star. However, the shortest orbital distance inferred to date for a directly-imaged planet is 3.53~au, for the massive exoplanet HD206893~c \citep{hinkley23}. Normalised this distance by the stellar radius of the main sequence F-type host star of 1.25~$R_{\sun}$ \citep{gaspar16}, this planet orbits at around 600~$R_\star$. An Alfv\'en surface of this size would require an incredibly strong magnetic field strength at the stellar surface. This would be unprecedented for a main sequence F-type star like HD206893, which typically exhibit large-scale surface field strengths of just a few Gauss \citep[e.g.][]{fares12, jeffers18, seach22}.

Another method which could be more feasible for detecting planets in C1 or C2 with current-generation telescopes is the astrometry method, which uses the reflex motion of the star projected on to the plane of the sky to infer the presence of a companion. This method is expected to lead to an explosion in the number of detected non-transiting exoplanets with survey telescopes such as Gaia \citep{perryman14, winn22}. To date, the shortest orbital distance planet discovered to orbit a main-sequence star via astrometry is the 2.3 Jupiter mass planet GJ~896Ab (EQ~Peg~Ab), which orbits its M3.5 host star at 0.639~au \citep{curiel22}. Interestingly, this detection was made using the Very Long Baseline Array (VLBA) at 8.4~GHz. Again normalising by the radius of the host star of 0.25~$R_{\sun}$, the planet orbits at 550 stellar radii. While the host star is an M~dwarf with an average large-scale surface magnetic field of around 500~G \citep{morin08}, it is unlikely its Alfv\'en surface extends to this distance. Nevertheless, a companion in a system similar to GJ~896A could be discovered using the same method if it is closer and more massive, such as a brown dwarf. Such systems would be very suitable candidates for discovery in tandem via magnetic SPI, as explored in this work.

It is therefore useful to also estimate what systems are most visible in the radio that we are also able to detect with current techniques, i.e. transiting exoplanets. For the planet to transit the stellar disk, its inclination must be in the range $|\cos i_\textrm{p}| < R_\star / a$, neglecting the radius of the planet. Using the same uniform distributions for each parameter as described in Section~\ref{sec:samples}, and limiting the values of $\cos i_\textrm{p}$ from $-R_\star / a$ to $R_\star / a$, we re-run our Monte Carlo sampling of the visibility function. 

For transiting systems, overall we find the same results as for systems where all orbital inclinations are considered, in which a randomly chosen system is visible for 4\% of the time, and 49\% of all systems are ever visible. Similarly, there is no preferential polarisation to the radio emission. This is unsurprising, given that the inclination of the rotation axis of the star and the magnetic obliquity, both of which remain unchanged in terms of sampling, are the dominant parameter as to whether radio emission is visible at all. It is not that transiting systems are not visible, but that planets in near face-on orbits result in maximum visibility (Figure~\ref{fig:c1-c2}). As such, the maximum duty cycle for transiting systems we find for our sample is 56\%, compared to 80\% for systems that cover the full $180\degr$ in orbital inclination. There are also no significant differences in terms of scatter plots of inclinations and magnetic obliquities that produce high duty cycles (Figure~\ref{fig:scatter star}) or the duty cycle against each parameter (Figure~\ref{fig:scatter plots}) for transiting systems.


\section{Summary \& Conclusions}

We now summarise and discuss the main findings of the paper.

\subsection{Current limitations to the model}

The model developed here, while fast and flexible, is not without limitations. The first of which is the lack of any information about the plasma itself. Accounting for such would allow for the plasma frequency to be computed, which provides a lower limit to the frequency range of emission \citep{kavanagh21, kavanagh22}, as well as the radio power emitted via the interaction to be estimated \citep[see][]{saur13}. It also allows for the calculation of the position of the Alfv\'en surface. Knowing this places a further constraint on the regions around the star where magnetic SPI can occur, without needing an arbitrary constraint like the maximum size of the field line as we have currently implemented. Knowledge of the plasma environment would also allow for absorption, reflection, and refraction processes to be incorporated into the radiative transfer. Accounting for these effects could significantly alter the visibility of the ECM emission. The plasma information also allows for the radio flux densities to be estimated, which in turn provides temporal modulation to the signal when visible. With this we can also compute the size of the planet's magnetosphere, which in turn influences the induced flux densities \citep{kavanagh22}.

Another aspect lacking from both the model presented here as well as that developed by \citet{kavanagh22} is the velocity distribution of electrons along the field line that is producing radio emission. Knowing this would allow us to self-consistently calculate many desirable quantities such as the frequency range, cone properties (which we assume to be constant in this work), and emission duration. However, for this one would likely have to couple an MHD simulation to a particle-in-cell type simulation, and account for how Alfv\'en waves are generated and propagate in a dynamical environment such as a stellar wind. Such a task is well beyond the scope of this paper, as well as in the case of MHD simulations such as those presented by \citet{kavanagh21} and \citet{kavanagh22}. However, it is still worth mentioning with future work in mind. Accounting for the evolution of the electron velocity distribution on the emitting field line would also allow us to compute properties such as the delay time between the interaction and the emission to appear, as well as trailing features such as those seen on in Io's footpoint on Jupiter in the UV \citep{hess10}. 

MHD models also provide information about the plasma inertia, which results in a toroidal component to the large-scale magnetic field that trails behind the direction of rotation in a Parker spiral-like configuration. Similarly, the closed field lines will be stretched outward radially, opening at some distance close to the Alfv\'en surface. Depending on the conditions, these deviations from purely dipolar field lines could alter the visibility of the emission. The same argument can be made against using a purely dipolar magnetic field as we have done in this work. If the induced emission is generated sufficiently close to the stellar surface, higher order modes of the magnetic field such as the quadrupole and octupolar will become more significant. As a result, the total magnetic field vector may deviate from the dipolar component significantly. However, knowledge of the strength of each magnetic mode is only generally obtainable via the ZDI method \cite[see also][]{lehmann22}. In future, it could be useful to also parameterise over the higher order magnetic modes in this context, using for example the potential field source surface method \citep[PFSSM][]{jardine02}. This however will be more numerically taxing than the assumption of a purely dipolar field.

In the future, it will also be necessary to develop numerical models for predicting the visibility of radio emission from magnetised low-mass stars that do not invoke the presence of a planet \citep[e.g.][]{llama18, owocki22}. Such models should also account for the aforementioned aspects mentioned in this Section such as propagation effects. Since the same underlying geometric calculations presented in this work are relevant in that regard, the MASER code can be adapted for these scenarios. Then in the case that emission is detected from a system, these models can ideally be utilised to uniquely identify the underlying generation mechanism \citep[see also][]{kavanagh22}.


\subsection{Comparison to the ExPRES code}
\label{sec:expres discusson}

It is worth noting that there are similarities between the code developed here and the ExPRES code developed by \citet{hess11} \citep[see also][]{louis19}. The key distinction is that ExPRES requires the pre-computed magnetic field geometry of the system as an input, whereas we compute the geometry of the field line the planet interacts with for an arbitrary set of system parameters on the fly. ExPRES also takes the plasma and energy of the electrons as inputs, which are used to prescribe the underlying electron cyclotron maser conditions. However, these conditions are highly uncertain in a stellar context, and likely require both MHD and particle-in-cell simulations to determine. Our code however does not explicitly assume that the prescriptions which appear to work well for the auroral emission on Jupiter and Saturn apply. ExPRES also is written in IDL, which is not open source. It is also unclear if it can be easily deployed for parametric studies, as we exhibit in this work with the MASER code.


\subsection{Concluding remarks}

In this work, we have developed a freely-available tool to assess and predict signatures of magnetic star-planet interactions in the radio regime. It is based on a key set of physical and geometrical parameters, which are generally known in part for exoplanetary systems. For systems with unknown parameters (i.e. the orbital distance of the planet), the model can be utilised in parametric studies to compare to observations that are indicative of such interactions. It is also fast, computationally inexpensive, and has low dependencies, and captures most of the key processes of the model presented in \citet{kavanagh22}, without the need for MHD simulations or magnetic field maps.

We first illustrated its ability to explain the phenomenon of radio emission appearing at the quadrature points of a satellite's orbit, which correspond to orbital phases of 0.25 and 0.75. However, this is in fact only possible in the case that the rotation, magnetic, and orbital axes are all aligned and lie in the plane of the sky. This is not the case in general for exoplanetary systems. Therefore, scheduling radio observations to coincide with the quadrature points of a known planetary orbit can result in the majority of the induced emission being missed.

We then utilised the model in a Monte Carlo simulation to assess which (if any) of the model parameters reflect exoplanetary systems we are biased towards detecting. Sampling the parameter space with 1 million values, we find that there are two distinct configurations where emission can be seen for up to 80\% of the time. This is significantly higher than the average value for systems that can ever be visible of $\sim9\%$. These two configurations rely on the inclination of the magnetic axis relative to the line of sight being fixed at an angle equal to that of the opening angle of the emission cone. Such configurations are possible if the magnetic and rotation axes are aligned (C1), or if we see the star pole on with an obliquity close to $90\degr$ (C2).

For C1, many M~dwarfs exhibit strong axisymmetric dipolar magnetic fields at their surfaces \citep{morin08, morin10}, and as a result they could be well-suited for detection of radio emission induced by planets in face-on orbits. For C2 however, it is not clear whether any M~dwarfs that have had their surface fields mapped with ZDI exhibit obliquities close to $90\degr$. Some early and late M~dwarfs do exhibit significant non-axisymmetric components, which in theory includes topologies with large obliquities. However, specific information relating to the dipolar component of the recovered magnetic field is often limited in the literature. Another interesting point is that if the magnetic field of the star evolves such that the dipole axis moves in to one of these configurations, emission may become more visible compared to other stages of the magnetic cycle. AD~Leo is an M~dwarf that has exhibited hints of activity cycles \citep{lavail18}, however, we have yet to see any evidence for a significant change to the dipole tilt.

In terms of the planet's orbital characteristics, we find that the most visible systems are those where the planet orbits over the magnetic poles. Combining this with the configurations C1 and C2 described above, these planets are in near face-on configurations. This is quite interesting, as such a population of exoplanets remains largely undiscovered via traditional methods, due to both their low radial velocity signatures and non-transiting nature. This could explain why none of the stars detected at radio wavelengths by \citet{callingham21b} are known to host any close-in planets. If that is the case, the astrometry method may prove to be very complementary for confirming their presence (see Section~\ref{sec:transiting}). We note that transiting exoplanets are still likely to be detectable when the star is in C1 or C2, but are less likely to be seen in blind radio surveys compared to planets in near face-on orbits. We also note that these results are based on our assumption of non-informative priors for the underlying system geometry. Further understanding of their true underlying distributions could alter these results.

Although the code developed here has been primarily discussed in a star-planet context, it can be easily adapted to any magnetised host-satellite system by simply changing the units of the input parameters (e.g. Section~\ref{sec:io}). In that sense, it may also be useful in future for interpreting radio emission from brown dwarfs and exoplanets. It also could be easily applied in the area of enhanced chromospheric/coronal emission from stars due to magnetic SPI \citep[e.g.][]{shkolnik03, lanza09, klein22}.


\section*{Acknowledgements}

We thank the anonymous reviewer for their helpful comments and suggestions. We acknowledge funding from the Dutch Research Council (NWO) for the e-MAPS (exploring magnetism on the planetary scale) project (project number VI.Vidi.203.093) under the NWO talent scheme Vidi. RDK also acknowledges funding from the European Research Council (ERC) under the European Union's Horizon 2020 research and innovation programme (grant agreement No. 817540, ASTROFLOW). We would also like to thank Benjamin Pope, Aline Vidotto, and Joan Bautista Climent for their insightful comments and suggestions on the manuscript.


\section*{Data availability}

All data presented in this work was generated using the \textsc{MASER} Python code we developed, which can is freely available on GitHub (see start of Section~\ref{sec:model}). We kindly request that utilising of this code in future publications acknowledge this work.


\bibliographystyle{mnras}
\bibliography{bibliography}

\begin{thebibliography}{}
\makeatletter
\relax
\def\mn@urlcharsother{\let\do\@makeother \do\$\do\&\do\#\do\^\do\_\do\%\do\~}
\def\mn@doi{\begingroup\mn@urlcharsother \@ifnextchar [ {\mn@doi@}
  {\mn@doi@[]}}
\def\mn@doi@[#1]#2{\def\@tempa{#1}\ifx\@tempa\@empty \href
  {http://dx.doi.org/#2} {doi:#2}\else \href {http://dx.doi.org/#2} {#1}\fi
  \endgroup}
\def\mn@eprint#1#2{\mn@eprint@#1:#2::\@nil}
\def\mn@eprint@arXiv#1{\href {http://arxiv.org/abs/#1} {{\tt arXiv:#1}}}
\def\mn@eprint@dblp#1{\href {http://dblp.uni-trier.de/rec/bibtex/#1.xml}
  {dblp:#1}}
\def\mn@eprint@#1:#2:#3:#4\@nil{\def\@tempa {#1}\def\@tempb {#2}\def\@tempc
  {#3}\ifx \@tempc \@empty \let \@tempc \@tempb \let \@tempb \@tempa \fi \ifx
  \@tempb \@empty \def\@tempb {arXiv}\fi \@ifundefined
  {mn@eprint@\@tempb}{\@tempb:\@tempc}{\expandafter \expandafter \csname
  mn@eprint@\@tempb\endcsname \expandafter{\@tempc}}}

\bibitem[\protect\citeauthoryear{{Albrecht}, {Dawson}  \& {Winn}}{{Albrecht}
  et~al.}{2022}]{albrecht22}
{Albrecht} S.~H.,  {Dawson} R.~I.,   {Winn} J.~N.,  2022, \mn@doi [\pasp]
  {10.1088/1538-3873/ac6c09}, \href
  {https://ui.adsabs.harvard.edu/abs/2022PASP..134h2001A} {134, 082001}

\bibitem[\protect\citeauthoryear{{Alfv{\'e}n}}{{Alfv{\'e}n}}{1942}]{alfven42}
{Alfv{\'e}n} H.,  1942, \mn@doi [\nat] {10.1038/150405d0}, \href
  {https://ui.adsabs.harvard.edu/abs/1942Natur.150..405A} {150, 405}

\bibitem[\protect\citeauthoryear{{Ashtari}, {Sciola}, {Turner}  \&
  {Stevenson}}{{Ashtari} et~al.}{2022}]{ashtari22}
{Ashtari} R.,  {Sciola} A.,  {Turner} J.~D.,   {Stevenson} K.,  2022, \mn@doi
  [\apj] {10.3847/1538-4357/ac92f5}, \href
  {https://ui.adsabs.harvard.edu/abs/2022ApJ...939...24A} {939, 24}

\bibitem[\protect\citeauthoryear{{Bagenal}}{{Bagenal}}{2013}]{bagenal13}
{Bagenal} F.,  2013, in {Oswalt} T.~D.,  {French} L.~M.,   {Kalas} P.,  eds, ,
  Planets, Stars and Stellar Systems. Volume 3: Solar and Stellar Planetary
  Systems.
p.~251, \mn@doi{10.1007/978-94-007-5606-9_6}

\bibitem[\protect\citeauthoryear{{Bagenal} \& {Dols}}{{Bagenal} \&
  {Dols}}{2020}]{bagenal20}
{Bagenal} F.,  {Dols} V.,  2020, \mn@doi [Journal of Geophysical Research
  (Space Physics)] {10.1029/2019JA027485}, \href
  {https://ui.adsabs.harvard.edu/abs/2020JGRA..12527485B} {125, e27485}

\bibitem[\protect\citeauthoryear{{Bigg}}{{Bigg}}{1964}]{bigg64}
{Bigg} E.~K.,  1964, \mn@doi [\nat] {10.1038/2031008a0}, \href
  {https://ui.adsabs.harvard.edu/abs/1964Natur.203.1008B} {203, 1008}

\bibitem[\protect\citeauthoryear{{Bills} \& {Scott}}{{Bills} \&
  {Scott}}{2022}]{bills22}
{Bills} B.~G.,  {Scott} B.~R.,  2022, \mn@doi [\planss]
  {10.1016/j.pss.2022.105474}, \href
  {https://ui.adsabs.harvard.edu/abs/2022P&SS..21905474B} {219, 105474}

\bibitem[\protect\citeauthoryear{{Blanco-Pozo} et~al.,}{{Blanco-Pozo}
  et~al.}{2023}]{blanco-pozo23}
{Blanco-Pozo} J.,  et~al., 2023, \mn@doi [\aap] {10.1051/0004-6361/202245053},
  \href {https://ui.adsabs.harvard.edu/abs/2023A&A...671A..50B} {671, A50}

\bibitem[\protect\citeauthoryear{{Burn}, {Schlecker}, {Mordasini},
  {Emsenhuber}, {Alibert}, {Henning}, {Klahr}  \& {Benz}}{{Burn}
  et~al.}{2021}]{burn21}
{Burn} R.,  {Schlecker} M.,  {Mordasini} C.,  {Emsenhuber} A.,  {Alibert} Y.,
  {Henning} T.,  {Klahr} H.,   {Benz} W.,  2021, \mn@doi [\aap]
  {10.1051/0004-6361/202140390}, \href
  {https://ui.adsabs.harvard.edu/abs/2021A&A...656A..72B} {656, A72}

\bibitem[\protect\citeauthoryear{{Callingham} et~al.,}{{Callingham}
  et~al.}{2021}]{callingham21b}
{Callingham} J.~R.,  et~al., 2021, \mn@doi [Nature Astronomy]
  {10.1038/s41550-021-01483-0}, \href
  {https://ui.adsabs.harvard.edu/abs/2021NatAs...5.1233C} {5, 1233}

\bibitem[\protect\citeauthoryear{{Connerney} et~al.,}{{Connerney}
  et~al.}{2022}]{connerney22}
{Connerney} J.~E.~P.,  et~al., 2022, \mn@doi [Journal of Geophysical Research
  (Planets)] {10.1029/2021JE007055}, \href
  {https://ui.adsabs.harvard.edu/abs/2022JGRE..12707055C} {127, e07055}

\bibitem[\protect\citeauthoryear{{Curiel}, {Ortiz-Le{\'o}n}, {Mioduszewski}  \&
  {Sanchez-Bermudez}}{{Curiel} et~al.}{2022}]{curiel22}
{Curiel} S.,  {Ortiz-Le{\'o}n} G.~N.,  {Mioduszewski} A.~J.,
  {Sanchez-Bermudez} J.,  2022, \mn@doi [\aj] {10.3847/1538-3881/ac7c66}, \href
  {https://ui.adsabs.harvard.edu/abs/2022AJ....164...93C} {164, 93}

\bibitem[\protect\citeauthoryear{{Das} \& {Chandra}}{{Das} \&
  {Chandra}}{2021}]{das21}
{Das} B.,  {Chandra} P.,  2021, \mn@doi [\apj] {10.3847/1538-4357/ac1075},
  \href {https://ui.adsabs.harvard.edu/abs/2021ApJ...921....9D} {921, 9}

\bibitem[\protect\citeauthoryear{{Donati} \& {Landstreet}}{{Donati} \&
  {Landstreet}}{2009}]{donati09}
{Donati} J.~F.,  {Landstreet} J.~D.,  2009, \mn@doi [\araa]
  {10.1146/annurev-astro-082708-101833}, \href
  {https://ui.adsabs.harvard.edu/abs/2009ARA&A..47..333D} {47, 333}

\bibitem[\protect\citeauthoryear{{Donati} et~al.,}{{Donati}
  et~al.}{2008}]{donati08}
{Donati} J.~F.,  et~al., 2008, \mn@doi [\mnras]
  {10.1111/j.1365-2966.2008.13799.x}, \href
  {https://ui.adsabs.harvard.edu/abs/2008MNRAS.390..545D} {390, 545}

\bibitem[\protect\citeauthoryear{{Drell}, {Foley}  \& {Ruderman}}{{Drell}
  et~al.}{1965}]{drell65}
{Drell} S.~D.,  {Foley} H.~M.,   {Ruderman} M.~A.,  1965, \mn@doi [\jgr]
  {10.1029/JZ070i013p03131}, \href
  {https://ui.adsabs.harvard.edu/abs/1965JGR....70.3131D} {70, 3131}

\bibitem[\protect\citeauthoryear{{Dulk}}{{Dulk}}{1985}]{dulk85}
{Dulk} G.~A.,  1985, \mn@doi [\araa] {10.1146/annurev.aa.23.090185.001125},
  \href {https://ui.adsabs.harvard.edu/abs/1985ARA&A..23..169D} {23, 169}

\bibitem[\protect\citeauthoryear{{Edler}, {de Gasperin}  \& {Rafferty}}{{Edler}
  et~al.}{2021}]{edler21}
{Edler} H.~W.,  {de Gasperin} F.,   {Rafferty} D.,  2021, \mn@doi [\aap]
  {10.1051/0004-6361/202140465}, \href
  {https://ui.adsabs.harvard.edu/abs/2021A&A...652A..37E} {652, A37}

\bibitem[\protect\citeauthoryear{{Fares} et~al.,}{{Fares}
  et~al.}{2012}]{fares12}
{Fares} R.,  et~al., 2012, \mn@doi [\mnras] {10.1111/j.1365-2966.2012.20780.x},
  \href {https://ui.adsabs.harvard.edu/abs/2012MNRAS.423.1006F} {423, 1006}

\bibitem[\protect\citeauthoryear{{G{\'a}sp{\'a}r}, {Rieke}  \&
  {Ballering}}{{G{\'a}sp{\'a}r} et~al.}{2016}]{gaspar16}
{G{\'a}sp{\'a}r} A.,  {Rieke} G.~H.,   {Ballering} N.,  2016, \mn@doi [\apj]
  {10.3847/0004-637X/826/2/171}, \href
  {https://ui.adsabs.harvard.edu/abs/2016ApJ...826..171G} {826, 171}

\bibitem[\protect\citeauthoryear{Harris et~al.,}{Harris et~al.}{2020}]{numpy}
Harris C.~R.,  et~al., 2020, \mn@doi [Nature] {10.1038/s41586-020-2649-2}, 585,
  357

\bibitem[\protect\citeauthoryear{{Hess} \& {Zarka}}{{Hess} \&
  {Zarka}}{2011}]{hess11}
{Hess} S.~L.~G.,  {Zarka} P.,  2011, \mn@doi [\aap]
  {10.1051/0004-6361/201116510}, \href
  {https://ui.adsabs.harvard.edu/abs/2011A&A...531A..29H} {531, A29}

\bibitem[\protect\citeauthoryear{{Hess}, {Delamere}, {Dols}, {Bonfond}  \&
  {Swift}}{{Hess} et~al.}{2010}]{hess10}
{Hess} S.~L.~G.,  {Delamere} P.,  {Dols} V.,  {Bonfond} B.,   {Swift} D.,
  2010, \mn@doi [Journal of Geophysical Research (Space Physics)]
  {10.1029/2009JA014928}, \href
  {https://ui.adsabs.harvard.edu/abs/2010JGRA..115.6205H} {115, A06205}

\bibitem[\protect\citeauthoryear{{Hinkley} et~al.,}{{Hinkley}
  et~al.}{2023}]{hinkley23}
{Hinkley} S.,  et~al., 2023, \mn@doi [\aap] {10.1051/0004-6361/202244727},
  \href {https://ui.adsabs.harvard.edu/abs/2023A&A...671L...5H} {671, L5}

\bibitem[\protect\citeauthoryear{{Jardine}, {Collier Cameron}  \&
  {Donati}}{{Jardine} et~al.}{2002}]{jardine02}
{Jardine} M.,  {Collier Cameron} A.,   {Donati} J.~F.,  2002, \mn@doi [\mnras]
  {10.1046/j.1365-8711.2002.05394.x}, \href
  {https://ui.adsabs.harvard.edu/abs/2002MNRAS.333..339J} {333, 339}

\bibitem[\protect\citeauthoryear{{Jeffers} et~al.,}{{Jeffers}
  et~al.}{2018}]{jeffers18}
{Jeffers} S.~V.,  et~al., 2018, \mn@doi [\mnras] {10.1093/mnras/sty1717}, \href
  {https://ui.adsabs.harvard.edu/abs/2018MNRAS.479.5266J} {479, 5266}

\bibitem[\protect\citeauthoryear{{Kavanagh} et~al.,}{{Kavanagh}
  et~al.}{2019}]{kavanagh19}
{Kavanagh} R.~D.,  et~al., 2019, \mn@doi [\mnras] {10.1093/mnras/stz655}, \href
  {https://ui.adsabs.harvard.edu/abs/2019MNRAS.485.4529K} {485, 4529}

\bibitem[\protect\citeauthoryear{{Kavanagh}, {Vidotto}, {Klein}, {Jardine},
  {Donati}  \& {{\'O} Fionnag{\'a}in}}{{Kavanagh} et~al.}{2021}]{kavanagh21}
{Kavanagh} R.~D.,  {Vidotto} A.~A.,  {Klein} B.,  {Jardine} M.~M.,  {Donati}
  J.-F.,   {{\'O} Fionnag{\'a}in} D.,  2021, \mn@doi [\mnras]
  {10.1093/mnras/stab929}, \href
  {https://ui.adsabs.harvard.edu/abs/2021MNRAS.504.1511K} {504, 1511}

\bibitem[\protect\citeauthoryear{{Kavanagh}, {Vidotto}, {Vedantham}, {Jardine},
  {Callingham}  \& {Morin}}{{Kavanagh} et~al.}{2022}]{kavanagh22}
{Kavanagh} R.~D.,  {Vidotto} A.~A.,  {Vedantham} H.~K.,  {Jardine} M.~M.,
  {Callingham} J.~R.,   {Morin} J.,  2022, \mn@doi [\mnras]
  {10.1093/mnras/stac1264}, \href
  {https://ui.adsabs.harvard.edu/abs/2022MNRAS.514..675K} {514, 675}

\bibitem[\protect\citeauthoryear{{Kivelson} \& {Russell}}{{Kivelson} \&
  {Russell}}{1995}]{kivelson95}
{Kivelson} M.~G.,  {Russell} C.~T.,  1995, {Introduction to Space Physics}, 1st
  edn.
Cambridge University Press, Cambridge, United Kingdom

\bibitem[\protect\citeauthoryear{{Klein} et~al.,}{{Klein}
  et~al.}{2022}]{klein22}
{Klein} B.,  et~al., 2022, \mn@doi [\mnras] {10.1093/mnras/stac761}, \href
  {https://ui.adsabs.harvard.edu/abs/2022MNRAS.512.5067K} {512, 5067}

\bibitem[\protect\citeauthoryear{{Kochukhov}}{{Kochukhov}}{2021}]{kochukhov21}
{Kochukhov} O.,  2021, \mn@doi [\aapr] {10.1007/s00159-020-00130-3}, \href
  {https://ui.adsabs.harvard.edu/abs/2021A&ARv..29....1K} {29, 1}

\bibitem[\protect\citeauthoryear{Lam, Pitrou  \& Seibert}{Lam
  et~al.}{2015}]{numba}
Lam S.~K.,  Pitrou A.,   Seibert S.,  2015, in Proceedings of the Second
  Workshop on the LLVM Compiler Infrastructure in HPC. LLVM '15.
Association for Computing Machinery, New York, NY, USA,
  \mn@doi{10.1145/2833157.2833162}, \url
  {https://doi.org/10.1145/2833157.2833162}

\bibitem[\protect\citeauthoryear{{Lamy} et~al.,}{{Lamy} et~al.}{2022}]{lamy22}
{Lamy} L.,  et~al., 2022, \mn@doi [Journal of Geophysical Research (Space
  Physics)] {10.1029/2021JA030160}, \href
  {https://ui.adsabs.harvard.edu/abs/2022JGRA..12730160L} {127, e30160}

\bibitem[\protect\citeauthoryear{{Lanza}}{{Lanza}}{2009}]{lanza09}
{Lanza} A.~F.,  2009, \mn@doi [\aap] {10.1051/0004-6361/200912367}, \href
  {https://ui.adsabs.harvard.edu/abs/2009A&A...505..339L} {505, 339}

\bibitem[\protect\citeauthoryear{{Lavail}, {Kochukhov}  \& {Wade}}{{Lavail}
  et~al.}{2018}]{lavail18}
{Lavail} A.,  {Kochukhov} O.,   {Wade} G.~A.,  2018, \mn@doi [\mnras]
  {10.1093/mnras/sty1825}, \href
  {https://ui.adsabs.harvard.edu/abs/2018MNRAS.479.4836L} {479, 4836}

\bibitem[\protect\citeauthoryear{{Lehmann} \& {Donati}}{{Lehmann} \&
  {Donati}}{2022}]{lehmann22}
{Lehmann} L.~T.,  {Donati} J.~F.,  2022, \mn@doi [\mnras]
  {10.1093/mnras/stac1519}, \href
  {https://ui.adsabs.harvard.edu/abs/2022MNRAS.514.2333L} {514, 2333}

\bibitem[\protect\citeauthoryear{{Lehmann}, {Hussain}, {Jardine}, {Mackay}  \&
  {Vidotto}}{{Lehmann} et~al.}{2019}]{lehmann19}
{Lehmann} L.~T.,  {Hussain} G.~A.~J.,  {Jardine} M.~M.,  {Mackay} D.~H.,
  {Vidotto} A.~A.,  2019, \mn@doi [\mnras] {10.1093/mnras/sty3362}, \href
  {https://ui.adsabs.harvard.edu/abs/2019MNRAS.483.5246L} {483, 5246}

\bibitem[\protect\citeauthoryear{{Lehmann}, {Hussain}, {Vidotto}, {Jardine}  \&
  {Mackay}}{{Lehmann} et~al.}{2021}]{lehmann21}
{Lehmann} L.~T.,  {Hussain} G.~A.~J.,  {Vidotto} A.~A.,  {Jardine} M.~M.,
  {Mackay} D.~H.,  2021, \mn@doi [\mnras] {10.1093/mnras/staa3284}, \href
  {https://ui.adsabs.harvard.edu/abs/2021MNRAS.500.1243L} {500, 1243}

\bibitem[\protect\citeauthoryear{{Llama}, {Jardine}, {Wood}, {Hallinan}  \&
  {Morin}}{{Llama} et~al.}{2018}]{llama18}
{Llama} J.,  {Jardine} M.~M.,  {Wood} K.,  {Hallinan} G.,   {Morin} J.,  2018,
  \mn@doi [\apj] {10.3847/1538-4357/aaa59f}, \href
  {https://ui.adsabs.harvard.edu/abs/2018ApJ...854....7L} {854, 7}

\bibitem[\protect\citeauthoryear{{Louis}, {Hess}, {Cecconi}, {Zarka}, {Lamy},
  {Aicardi}  \& {Loh}}{{Louis} et~al.}{2019}]{louis19}
{Louis} C.~K.,  {Hess} S.~L.~G.,  {Cecconi} B.,  {Zarka} P.,  {Lamy} L.,
  {Aicardi} S.,   {Loh} A.,  2019, \mn@doi [\aap]
  {10.1051/0004-6361/201935161}, \href
  {https://ui.adsabs.harvard.edu/abs/2019A&A...627A..30L} {627, A30}

\bibitem[\protect\citeauthoryear{{Lu}, {Curtis}, {Angus}, {David}  \&
  {Hattori}}{{Lu} et~al.}{2022}]{lu22}
{Lu} Y.~L.,  {Curtis} J.~L.,  {Angus} R.,  {David} T.~J.,   {Hattori} S.,
  2022, \mn@doi [\aj] {10.3847/1538-3881/ac9bee}, \href
  {https://ui.adsabs.harvard.edu/abs/2022AJ....164..251L} {164, 251}

\bibitem[\protect\citeauthoryear{{Marques}, {Zarka}, {Echer}, {Ryabov},
  {Alves}, {Denis}  \& {Coffre}}{{Marques} et~al.}{2017}]{marques17}
{Marques} M.~S.,  {Zarka} P.,  {Echer} E.,  {Ryabov} V.~B.,  {Alves} M.~V.,
  {Denis} L.,   {Coffre} A.,  2017, \mn@doi [\aap]
  {10.1051/0004-6361/201630025}, \href
  {https://ui.adsabs.harvard.edu/abs/2017A&A...604A..17M} {604, A17}

\bibitem[\protect\citeauthoryear{{Melrose} \& {Dulk}}{{Melrose} \&
  {Dulk}}{1982}]{melrose82}
{Melrose} D.~B.,  {Dulk} G.~A.,  1982, \mn@doi [\apj] {10.1086/160219}, \href
  {https://ui.adsabs.harvard.edu/abs/1982ApJ...259..844M} {259, 844}

\bibitem[\protect\citeauthoryear{{Morin} et~al.,}{{Morin}
  et~al.}{2008}]{morin08}
{Morin} J.,  et~al., 2008, \mn@doi [\mnras] {10.1111/j.1365-2966.2008.13809.x},
  \href {https://ui.adsabs.harvard.edu/abs/2008MNRAS.390..567M} {390, 567}

\bibitem[\protect\citeauthoryear{{Morin}, {Donati}, {Petit}, {Delfosse},
  {Forveille}  \& {Jardine}}{{Morin} et~al.}{2010}]{morin10}
{Morin} J.,  {Donati} J.~F.,  {Petit} P.,  {Delfosse} X.,  {Forveille} T.,
  {Jardine} M.~M.,  2010, \mn@doi [\mnras] {10.1111/j.1365-2966.2010.17101.x},
  \href {https://ui.adsabs.harvard.edu/abs/2010MNRAS.407.2269M} {407, 2269}

\bibitem[\protect\citeauthoryear{{Neubauer}}{{Neubauer}}{1980}]{neubauer80}
{Neubauer} F.~M.,  1980, \mn@doi [\jgr] {10.1029/JA085iA03p01171}, \href
  {https://ui.adsabs.harvard.edu/abs/1980JGR....85.1171N} {85, 1171}

\bibitem[\protect\citeauthoryear{{Nicholson}, {Parker}, {Church}, {Davies},
  {Fearon}  \& {Walton}}{{Nicholson} et~al.}{2019}]{nicholson19}
{Nicholson} R.~B.,  {Parker} R.~J.,  {Church} R.~P.,  {Davies} M.~B.,  {Fearon}
  N.~M.,   {Walton} S. R.~J.,  2019, \mn@doi [\mnras] {10.1093/mnras/stz606},
  \href {https://ui.adsabs.harvard.edu/abs/2019MNRAS.485.4893N} {485, 4893}

\bibitem[\protect\citeauthoryear{{Owocki}, {Shultz}, {ud-Doula}, {Chandra},
  {Das}  \& {Leto}}{{Owocki} et~al.}{2022}]{owocki22}
{Owocki} S.~P.,  {Shultz} M.~E.,  {ud-Doula} A.,  {Chandra} P.,  {Das} B.,
  {Leto} P.,  2022, \mn@doi [\mnras] {10.1093/mnras/stac341}, \href
  {https://ui.adsabs.harvard.edu/abs/2022MNRAS.513.1449O} {513, 1449}

\bibitem[\protect\citeauthoryear{{Panchenko} \& {Rucker}}{{Panchenko} \&
  {Rucker}}{2016}]{pachenko16}
{Panchenko} M.,  {Rucker} H.~O.,  2016, \mn@doi [\aap]
  {10.1051/0004-6361/201527397}, \href
  {https://ui.adsabs.harvard.edu/abs/2016A&A...596A..18P} {596, A18}

\bibitem[\protect\citeauthoryear{{P{\'e}rez-Torres} et~al.,}{{P{\'e}rez-Torres}
  et~al.}{2021}]{perez-torres21}
{P{\'e}rez-Torres} M.,  et~al., 2021, \mn@doi [\aap]
  {10.1051/0004-6361/202039052}, \href
  {https://ui.adsabs.harvard.edu/abs/2021A&A...645A..77P} {645, A77}

\bibitem[\protect\citeauthoryear{{Perryman}, {Hartman}, {Bakos}  \&
  {Lindegren}}{{Perryman} et~al.}{2014}]{perryman14}
{Perryman} M.,  {Hartman} J.,  {Bakos} G.~{\'A}.,   {Lindegren} L.,  2014,
  \mn@doi [\apj] {10.1088/0004-637X/797/1/14}, \href
  {https://ui.adsabs.harvard.edu/abs/2014ApJ...797...14P} {797, 14}

\bibitem[\protect\citeauthoryear{{Pineda} \& {Villadsen}}{{Pineda} \&
  {Villadsen}}{2023}]{pineda23}
{Pineda} J.~S.,  {Villadsen} J.,  2023, \mn@doi [Nature Astronomy]
  {10.1038/s41550-023-01914-0}, \href
  {https://ui.adsabs.harvard.edu/abs/2023NatAs.tmp...65P} {}

\bibitem[\protect\citeauthoryear{{Popinchalk}, {Faherty}, {Kiman}, {Gagn{\'e}},
  {Curtis}, {Angus}, {Cruz}  \& {Rice}}{{Popinchalk}
  et~al.}{2021}]{popinchalk21}
{Popinchalk} M.,  {Faherty} J.~K.,  {Kiman} R.,  {Gagn{\'e}} J.,  {Curtis}
  J.~L.,  {Angus} R.,  {Cruz} K.~L.,   {Rice} E.~L.,  2021, \mn@doi [\apj]
  {10.3847/1538-4357/ac0444}, \href
  {https://ui.adsabs.harvard.edu/abs/2021ApJ...916...77P} {916, 77}

\bibitem[\protect\citeauthoryear{{Queinnec} \& {Zarka}}{{Queinnec} \&
  {Zarka}}{1998}]{queinnec98}
{Queinnec} J.,  {Zarka} P.,  1998, \mn@doi [\jgr] {10.1029/98JA02435}, \href
  {https://ui.adsabs.harvard.edu/abs/1998JGR...10326649Q} {103, 26649}

\bibitem[\protect\citeauthoryear{{Rappaport}, {Sanchis-Ojeda}, {Rogers},
  {Levine}  \& {Winn}}{{Rappaport} et~al.}{2013}]{rappaport13}
{Rappaport} S.,  {Sanchis-Ojeda} R.,  {Rogers} L.~A.,  {Levine} A.,   {Winn}
  J.~N.,  2013, \mn@doi [\apjl] {10.1088/2041-8205/773/1/L15}, \href
  {https://ui.adsabs.harvard.edu/abs/2013ApJ...773L..15R} {773, L15}

\bibitem[\protect\citeauthoryear{{Reinhold}, {Bell}, {Kuszlewicz}, {Hekker}  \&
  {Shapiro}}{{Reinhold} et~al.}{2019}]{reinhold19}
{Reinhold} T.,  {Bell} K.~J.,  {Kuszlewicz} J.,  {Hekker} S.,   {Shapiro}
  A.~I.,  2019, \mn@doi [\aap] {10.1051/0004-6361/201833754}, \href
  {https://ui.adsabs.harvard.edu/abs/2019A&A...621A..21R} {621, A21}

\bibitem[\protect\citeauthoryear{{Ribas} et~al.,}{{Ribas}
  et~al.}{2023}]{ribas23}
{Ribas} I.,  et~al., 2023, \mn@doi [\aap] {10.1051/0004-6361/202244879}, \href
  {https://ui.adsabs.harvard.edu/abs/2023A&A...670A.139R} {670, A139}

\bibitem[\protect\citeauthoryear{{Saur}, {Grambusch}, {Duling}, {Neubauer}  \&
  {Simon}}{{Saur} et~al.}{2013}]{saur13}
{Saur} J.,  {Grambusch} T.,  {Duling} S.,  {Neubauer} F.~M.,   {Simon} S.,
  2013, \mn@doi [\aap] {10.1051/0004-6361/201118179}, \href
  {https://ui.adsabs.harvard.edu/abs/2013A&A...552A.119S} {552, A119}

\bibitem[\protect\citeauthoryear{{Schlecker} et~al.,}{{Schlecker}
  et~al.}{2022}]{schlecker22}
{Schlecker} M.,  et~al., 2022, \mn@doi [\aap] {10.1051/0004-6361/202142543},
  \href {https://ui.adsabs.harvard.edu/abs/2022A&A...664A.180S} {664, A180}

\bibitem[\protect\citeauthoryear{{Schweitzer} et~al.,}{{Schweitzer}
  et~al.}{2019}]{schweitzer19}
{Schweitzer} A.,  et~al., 2019, \mn@doi [\aap] {10.1051/0004-6361/201834965},
  \href {https://ui.adsabs.harvard.edu/abs/2019A&A...625A..68S} {625, A68}

\bibitem[\protect\citeauthoryear{{Seach}, {Marsden}, {Carter}, {Evensberget},
  {Folsom}, {Neiner}  \& {Mengel}}{{Seach} et~al.}{2022}]{seach22}
{Seach} J.~M.,  {Marsden} S.~C.,  {Carter} B.~D.,  {Evensberget} D.,  {Folsom}
  C.~P.,  {Neiner} C.,   {Mengel} M.~W.,  2022, \mn@doi [\mnras]
  {10.1093/mnras/stab3289}, \href
  {https://ui.adsabs.harvard.edu/abs/2022MNRAS.509.5117S} {509, 5117}

\bibitem[\protect\citeauthoryear{{Shkolnik}, {Walker}  \&
  {Bohlender}}{{Shkolnik} et~al.}{2003}]{shkolnik03}
{Shkolnik} E.,  {Walker} G.~A.~H.,   {Bohlender} D.~A.,  2003, \mn@doi [\apj]
  {10.1086/378583}, \href
  {https://ui.adsabs.harvard.edu/abs/2003ApJ...597.1092S} {597, 1092}

\bibitem[\protect\citeauthoryear{{Shulyak}, {Reiners}, {Engeln}, {Malo},
  {Yadav}, {Morin}  \& {Kochukhov}}{{Shulyak} et~al.}{2017}]{shulyak17}
{Shulyak} D.,  {Reiners} A.,  {Engeln} A.,  {Malo} L.,  {Yadav} R.,  {Morin}
  J.,   {Kochukhov} O.,  2017, \mn@doi [Nature Astronomy]
  {10.1038/s41550-017-0184}, \href
  {https://ui.adsabs.harvard.edu/abs/2017NatAs...1E.184S} {1, 0184}

\bibitem[\protect\citeauthoryear{{Stef{\`a}nsson} et~al.,}{{Stef{\`a}nsson}
  et~al.}{2022}]{stefansson22}
{Stef{\`a}nsson} G.,  et~al., 2022, \mn@doi [\apjl] {10.3847/2041-8213/ac6e3c},
  \href {https://ui.adsabs.harvard.edu/abs/2022ApJ...931L..15S} {931, L15}

\bibitem[\protect\citeauthoryear{{Treumann}}{{Treumann}}{2006}]{treumann06}
{Treumann} R.~A.,  2006, \mn@doi [\aapr] {10.1007/s00159-006-0001-y}, \href
  {https://ui.adsabs.harvard.edu/abs/2006A&ARv..13..229T} {13, 229}

\bibitem[\protect\citeauthoryear{{Triaud}}{{Triaud}}{2018}]{triaud18}
{Triaud} A. H.~M.~J.,  2018, in {Deeg} H.~J.,  {Belmonte} J.~A.,  eds, ,
  Handbook of Exoplanets.
p.~2, \mn@doi{10.1007/978-3-319-55333-7_2}

\bibitem[\protect\citeauthoryear{{Trigilio} et~al.,}{{Trigilio}
  et~al.}{2023}]{trigilio23}
{Trigilio} C.,  et~al., 2023, \mn@doi [arXiv e-prints]
  {10.48550/arXiv.2305.00809}, \href
  {https://ui.adsabs.harvard.edu/abs/2023arXiv230500809T} {p. arXiv:2305.00809}

\bibitem[\protect\citeauthoryear{{Unterborn} \& {Panero}}{{Unterborn} \&
  {Panero}}{2019}]{unterborn19}
{Unterborn} C.~T.,  {Panero} W.~R.,  2019, \mn@doi [Journal of Geophysical
  Research (Planets)] {10.1029/2018JE005844}, \href
  {https://ui.adsabs.harvard.edu/abs/2019JGRE..124.1704U} {124, 1704}

\bibitem[\protect\citeauthoryear{{Vedantham} et~al.,}{{Vedantham}
  et~al.}{2020}]{vedantham20}
{Vedantham} H.~K.,  et~al., 2020, \mn@doi [Nature Astronomy]
  {10.1038/s41550-020-1011-9}, \href
  {https://ui.adsabs.harvard.edu/abs/2020NatAs...4..577V} {4, 577}

\bibitem[\protect\citeauthoryear{{Vidotto}, {Jardine}, {Morin}, {Donati},
  {Opher}  \& {Gombosi}}{{Vidotto} et~al.}{2014}]{vidotto14}
{Vidotto} A.~A.,  {Jardine} M.,  {Morin} J.,  {Donati} J.~F.,  {Opher} M.,
  {Gombosi} T.~I.,  2014, \mn@doi [\mnras] {10.1093/mnras/stt2265}, \href
  {https://ui.adsabs.harvard.edu/abs/2014MNRAS.438.1162V} {438, 1162}

\bibitem[\protect\citeauthoryear{{Villadsen} \& {Hallinan}}{{Villadsen} \&
  {Hallinan}}{2019}]{villadsen19}
{Villadsen} J.,  {Hallinan} G.,  2019, \mn@doi [\apj]
  {10.3847/1538-4357/aaf88e}, \href
  {https://ui.adsabs.harvard.edu/abs/2019ApJ...871..214V} {871, 214}

\bibitem[\protect\citeauthoryear{{Winn}}{{Winn}}{2022}]{winn22}
{Winn} J.~N.,  2022, \mn@doi [\aj] {10.3847/1538-3881/ac9126}, \href
  {https://ui.adsabs.harvard.edu/abs/2022AJ....164..196W} {164, 196}

\bibitem[\protect\citeauthoryear{{Winters} et~al.,}{{Winters}
  et~al.}{2019}]{winters19}
{Winters} J.~G.,  et~al., 2019, \mn@doi [\aj] {10.3847/1538-3881/ab05dc}, \href
  {https://ui.adsabs.harvard.edu/abs/2019AJ....157..216W} {157, 216}

\bibitem[\protect\citeauthoryear{{Zarka}}{{Zarka}}{2007}]{zarka07}
{Zarka} P.,  2007, \mn@doi [\planss] {10.1016/j.pss.2006.05.045}, \href
  {https://ui.adsabs.harvard.edu/abs/2007P&SS...55..598Z} {55, 598}

\makeatother
\end{thebibliography}


\appendix

\section{Vectors for the stellar rotation and magnetic axes}
\label{sec:star vectors}

\begin{figure}
\centering
\includegraphics[width = \columnwidth]{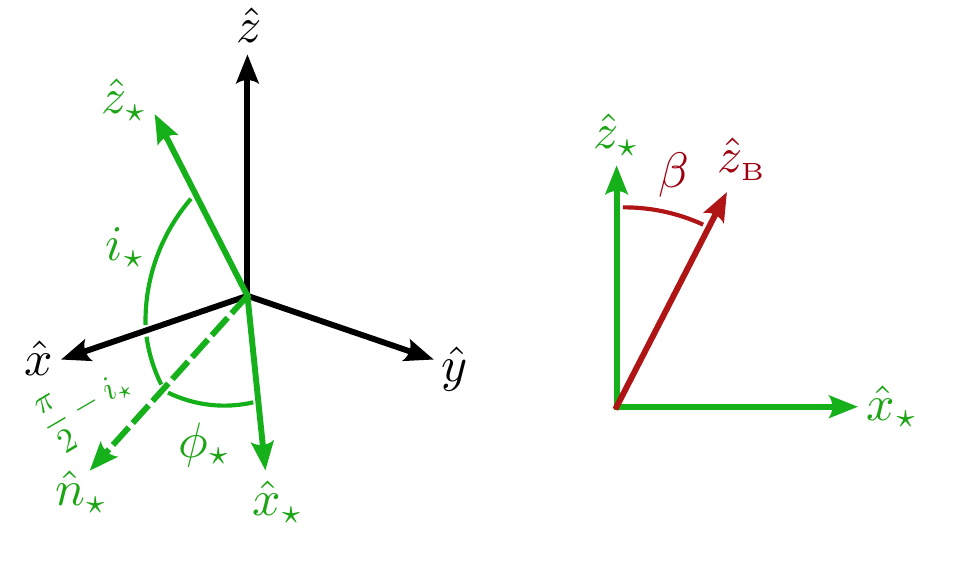}
\caption{Coordinate systems describing the rotation of the star about its axis $\hat{z}_\star$ (left) and the magnetic obliquity of the dipole axis $\hat{z}_\textrm{B}$ (right).}
\label{fig:star vectors}
\end{figure}

In this work, we relate all vectors describing the exoplanetary system to the line of sight vector $\hat{x} = (1, 0, 0)$, the projection of the stellar rotation axis $\hat{z}_\star$ onto the plane of the sky $\hat{z} = (0, 0, 1)$, and the vector perpendicular to $\hat{z}$ in the plane of the sky $\hat{y} = \hat{z}\times\hat{x} = (0, 1, 0)$. The rotation axis is inclined relative to the line of sight $\hat{x}$ by the angle $i_\star$:
\begin{equation}
\hat{z}_\star = \cos i_\star \hat{x} + \sin i_\star \hat{z} .
\label{eq:z_s}
\end{equation}
The magnetic axis $\hat{z}_\textrm{B}$ is tilted relative to the rotation axis by the angle $\beta$, and the projection of $\hat{z}_\textrm{B}$ on to the star's equatorial plane is $\hat{x}_\star$. The rotation phase of the star $\phi_\star$ is measured between $\hat{x}_\star$ and the vector $\hat{n}_\star$, which is the projection of $\hat{x}$ onto the equatorial plane:
\begin{equation}
\hat{n}_\star = \sin i_\star \hat{x} - \cos i_\star \hat{z} .
\end{equation}
The rotation phase $\phi_\star = 0$ when the $\hat{x}_\star = \hat{n}_\star$. The vector $\hat{x}_\star$ is therefore

\begin{equation}
\hat{x}_\star = \cos\phi_\star\hat{n}_\star + \sin\phi_\star\hat{y} ,
\end{equation}
and the magnetic axis is
\begin{equation}
\hat{z}_\textrm{B} = \sin\beta\hat{x}_\star + \cos\beta\hat{z}_\star .
\label{eq:z_B}
\end{equation}
Figure~\ref{fig:star vectors} shows a sketch of the vectors described here.

\section{Vectors for the planet position and spin-orbit misalignment}
\label{sec:planet vectors}

\begin{figure}
\centering
\includegraphics[width = \columnwidth]{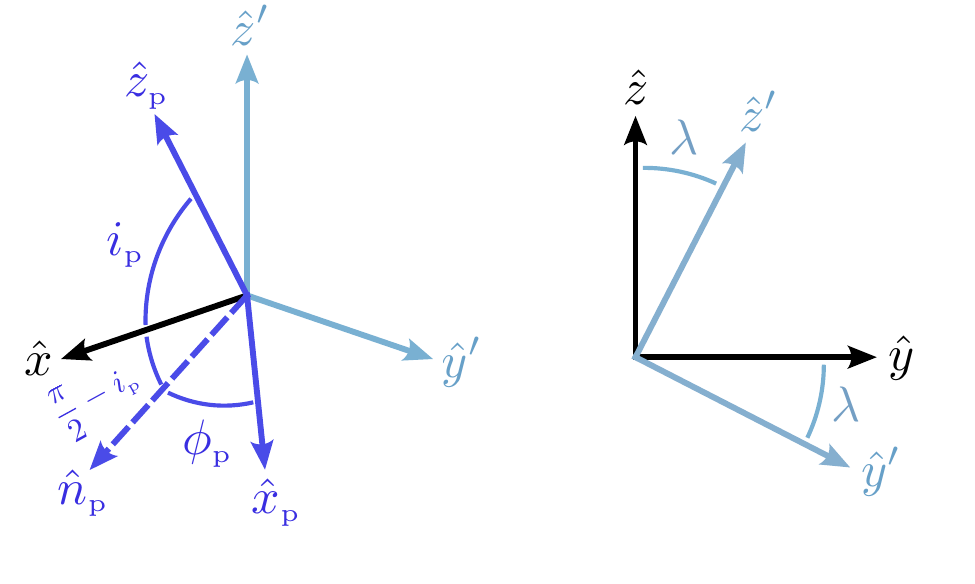}
\caption{Vectors describing the position of the planet $\hat{x}_\textrm{p}$ (left) and the spin-orbit misalignment (right), which is characterised by the angle $\lambda$, the angle formed by the projection of the rotation and orbital axes on to the plane of the sky.}
\label{fig:planet vectors}
\end{figure}

Around the star, a planet orbits. The normal to its orbital plane is $\hat{z}_\textrm{p}$, which is inclined relative to the line of sight by the angle $i_\textrm{p}$. The projection of $\hat{z}_\textrm{p}$ on to the plane of the sky is $\hat{z}'$:
\begin{equation}
\hat{z}_\textrm{p} = \cos i_\textrm{p} \hat{x} + \sin i_\textrm{p} \hat{z}' .
\label{eq:z_p}
\end{equation}
In general, $\hat{z}'$ is not aligned with $\hat{z}$, the projection of the stellar rotation axis on to the plane of the sky, and the angle measured from $\hat{z}'$ to $\hat{z}$ is $\lambda$. This is known as the \textit{projected} spin-orbit angle. Similarly, the angle from $\hat{y}$ to the vector perpendicular to $\hat{z}'$ in the plane of the sky $\hat{y}' = \hat{z}'\times\hat{x}$ is also $\lambda$. $\hat{y}'$ and $\hat{z}'$ can be expressed as
\begin{align}
& \hat{y}' = \cos\lambda\hat{y} - \sin\lambda\hat{z} , \\
& \hat{z}' = \sin\lambda\hat{y} + \cos\lambda\hat{z} .
\end{align}
The \textit{true} spin-orbit angle $\psi$ is the angle between the rotation and orbital axes, which is:
\begin{equation}
\cos\psi = \hat{z}_\star \cdot \hat{z}_\textrm{p} = \cos i_\star \cos i_\textrm{p} + \sin i_\star \sin i_\textrm{p} \cos\lambda .
\label{eq:spin orbit angle}
\end{equation}
The rotation phase of the planet is measured between the position of the planet $\hat{x}_\textrm{p}$ and the projection of the line of sight on to the orbital plane $\hat{n}_\textrm{p}$, which is
\begin{equation}
\hat{n}_\textrm{p} = \sin i_\textrm{p} \hat{x} - \cos i_\textrm{p} \hat{z}' .
\end{equation}
Therefore, the position of the planet is given by:
\begin{equation}
\hat{x}_\textrm{p} = \cos\phi_\textrm{p}\hat{n}_\textrm{p} + \sin\phi_\textrm{p}\hat{y}' .
\label{eq:x_p}
\end{equation}
A sketch of the vectors described here is shown in Figure~\ref{fig:planet vectors}.

\section{Finding the root of Equation~\ref{EQ:DIPOLAR FIELD LINE SOLVE}}
\label{sec:newthon method}

To find the root of Equation~\ref{EQ:DIPOLAR FIELD LINE SOLVE}, we use Newton's method, which utilises the derivative of the function to be solved. The derivative of Equation~\ref{EQ:DIPOLAR FIELD LINE SOLVE} with respect to $r$ is
\begin{equation}
F' = \frac{6}{R_\star} \Big(\frac{B}{B_\star}\Big)^2 \Big(\frac{r}{R_\star}\Big)^5 + \frac{3}{4L} .
\label{eq:newton method derivative}
\end{equation}
From an initial value of $r = r_i$, we linearly extrapolate the tangent line from the point $(r_i, F(r_i))$ to the point where $F = 0$. The value of $r$ where this line crosses $F = 0$ is $r_{i+1}$, which can be expressed as
\begin{equation}
r_{i+1} = r_i - \frac{F(r_i)}{F'(r_i)} .
\end{equation}
We iterate this process is until $|B_\nu - B(r_i)| / B_\nu$ is less than 1\%. Initialising the value of $r_i = R_\star$, this typically takes 10 to 50 iterations depending on the values of the coefficients of Equation~\ref{EQ:DIPOLAR FIELD LINE SOLVE}. At that point, we take the value of $r_\nu = r_i$, and then compute $\theta_\nu$ via Equation~\ref{eq:dipolar field radius}.


\section{The magnetic field vector along the field line}
\label{sec:field line vector}

\begin{figure}
\centering
\includegraphics[width = 0.5\columnwidth]{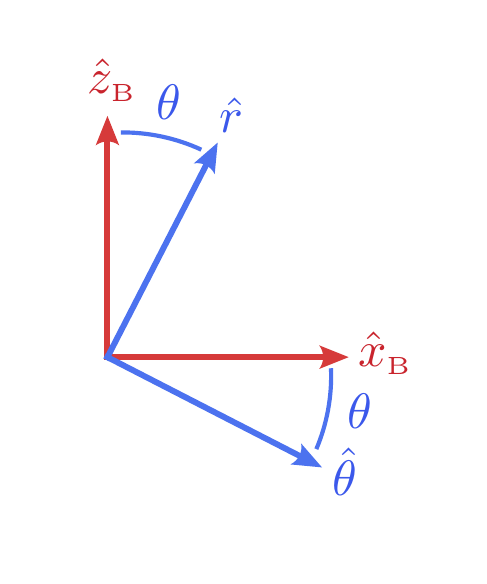}
\caption{The relation between the radial and meridional vectors at an emitting point at a co-latitude $\theta$ on the magnetic field line to the magnetic axis and equator of the star $\hat{z}_\textrm{B}$ and $\hat{x}_\textrm{B}$.}
\label{fig:emitting field vectors}
\end{figure}

At each point on the magnetic field line, the field vector can be decomposed into a radial and meridional (polar) component:
\begin{equation}
\vec{B} = B_r \hat{r} + B_\theta \hat{\theta} .
\label{eq:B vec}
\end{equation}
Here, $B_r$ and $B_\theta$ are the radial and meridional components, which at the point ($r$, $\theta$) are \citep{kivelson95}:
\begin{align}
& B_r = B_\star \Big(\frac{R_\star}{r}\Big)^3\cos\theta, \label{eq:Br} \\
& B_\theta = \frac{B_\star}{2} \Big(\frac{R_\star}{r}\Big)^3\sin\theta .
\end{align}
The radial and meridional unit vectors $\hat{r}$ and $\hat{\theta}$ can be expressed in terms of $\hat{x}_\textrm{B}$ and $\hat{z}_\textrm{B}$, which define the plane that the magnetic field line lies in (see Figures~\ref{fig:sketch field line} and \ref{fig:emitting field vectors}):
\begin{align}
& \hat{r} = \sin\theta\hat{x}_\textrm{B} + \cos\theta\hat{z}_\textrm{B}, \\
& \hat{\theta} = \cos\theta\hat{x}_\textrm{B} - \sin\theta\hat{z}_\textrm{B} . \label{eq:theta hat}
\end{align}
In the Northern magnetic hemisphere the emission cones point along $\hat{c}$, which are aligned with $\vec{B}$, i.e. $\hat{c} = \vec{B} / B$. In the Southern magnetic hemisphere however, the field lines point towards the stellar surface, but the emission cones are still oriented away from the surface. Therefore, in the Southern hemisphere, $\hat{c} = - \vec{B} / B$.

\section{Io-induced emission from Jupiter at quadrature}
\label{sec:io}

The Jupiter-Io interaction is a good example of the aligned scenario discussed in Section~\ref{sec:quadrature}. \citet{marques17} analysed 26 years of 10 -- 40~MHz radio data from Jupiter, identifying the components in its dynamic spectra that are due to the sub-Alfv\'enic interaction with Io. Naturally, this is a great dataset to benchmark our model against, replacing the star with Jupiter and the planet with Io. The relevant properties of Jupiter and Io are listed in Table~\ref{tab:jupiter io}. 

\begin{figure}
\centering
\includegraphics[width = \columnwidth]{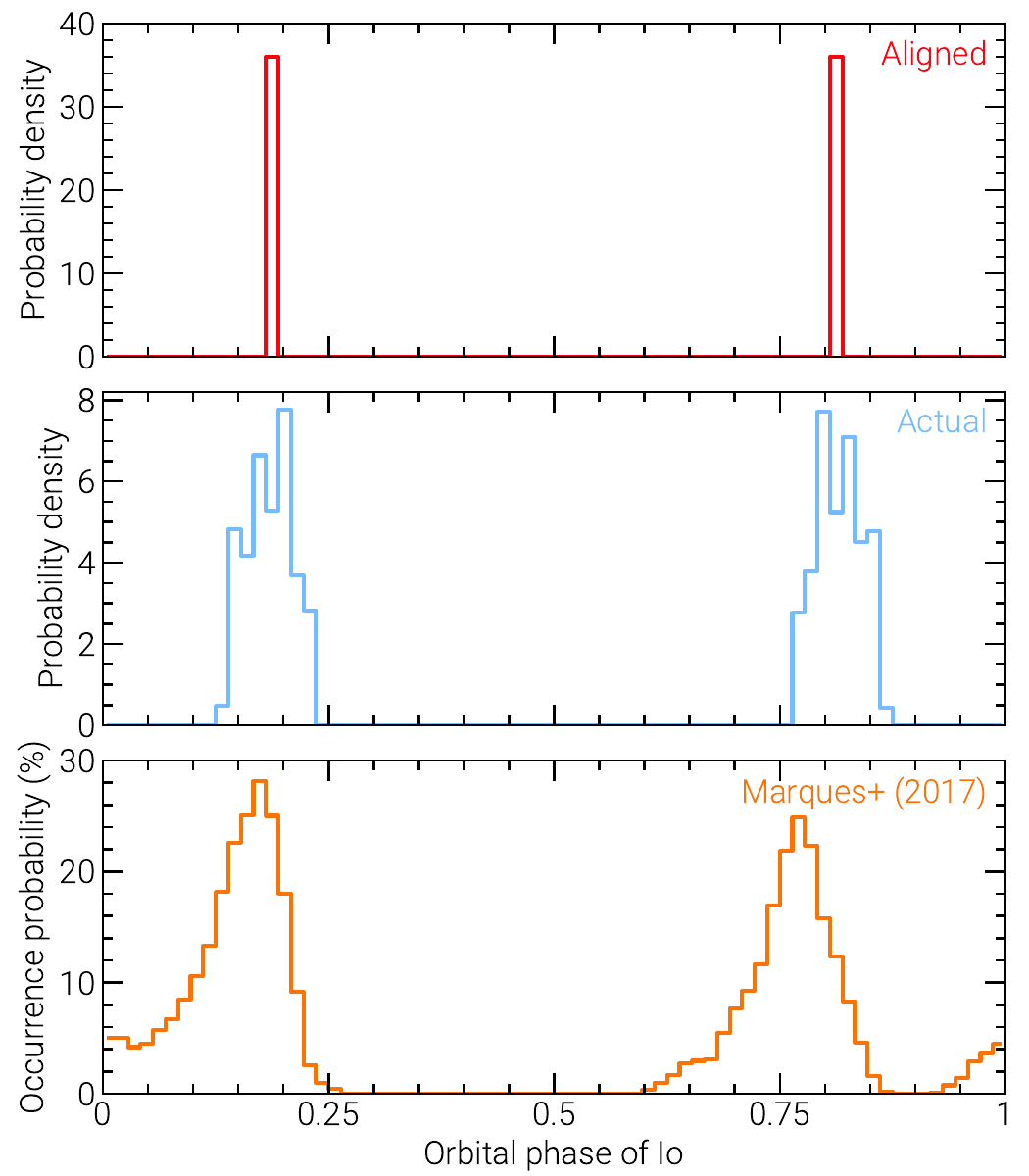}
\caption{Comparison of the probability density (PD) estimated for the Io-induced emission on Jupiter to the occurrence rate as a function of the orbital phase of Io inferred from 26 years of radio data by \citet{marques17}. The top panel shows the probability density for the Jupiter-Io system in an aligned configuration, the middle panel shows the PD for its actual configuration, and the bottom panel shows the occurrence probability from Figure 7(a) of \citet{marques17}. Note that phase zero in their figure corresponds to an orbital phase of 0.5 in the convention we adopt in this work (Section~\ref{sec:geometry}). Recall also that the assumed large-scale magnetic field is a dipole. Jupiter's surface magnetic field however exhibits higher order modes in addition to its dominant dipolar component \citep[see][]{connerney22}.}
\label{fig:io}
\end{figure}

In Figure~\ref{fig:io} we compare the results of \citet{marques17} to the PD of the lightcurve from a system described by the values listed in Table~\ref{tab:jupiter io}. For comparison we also show the PD of the lightcurve for the same system, but with the angles describing an aligned system (e.g.~Table~\ref{tab:quadrature values}). We compute both lightcurves for 500 orbits of Io, with 1000 time samples per orbit. We fix the initial rotation and orbital phases at zero. For the emission cone, we set the opening angle to $75\degr$ and thickness to $1\degr$ (Section~\ref{sec:radio}). For the observing frequency, we choose a value of 10~MHz.

We see a broadening of the probability density when the values deviate slightly from an aligned configuration. However, there are still two peaks centered about the orbital phases for the aligned case. The actual values reproduce a probability density that accurately resembles the long-term results of \citet{marques17}. Note that there is a slight discrepancy, in that the results of \citet{marques17} show that some emission occurs earlier on in Io's orbit, left of the two peaks. They attribute this to the fact that Jupiter rotates faster than Io's orbit. As the Alfv\'en waves have a finite velocity, by the time they interact with and accelerate the electrons that power the radio emission to near the surface, the field line has passed by Io. We do not account for such a phenomenon in our model however.

\begin{table}
\caption{Relevant parameters of Jupiter and Io}
\label{tab:jupiter io}
\centering
\begin{tabular}{lcc}
\hline
Parameter & Value & Reference \\
\hline
\underline{Jupiter:} \\
Mass & $1.90\times10^{30}$ g & \\
Radius & $7.15\times10^9$ cm & \\
Rotation period & 0.41 days & 1 \\
Inclination of rotation axis & 86.9$\degr$ & 1 \\
Dipole field strength (pole) & 8.6 G & 1 \\
Magnetic obliquity & $-9.4\degr$ & 1 \\
\underline{Io}: \\
Orbital distance & 5.90 Jupiter radii & 2 \\
Orbital inclination & $\sim86.9\degr$ & 3 \\
True spin-orbit angle & $\sim0\degr$ & 3 \\
\hline
\multicolumn{3}{p{0.8\columnwidth}}{1: \cite{bagenal13}; 2: \cite{bagenal20}; 3: The rotation axis of Jupiter and orbital plane of Io are separated by a very small angle \citep{bills22}.}
\end{tabular}
\end{table}

\section{Signal visibility as a function of each model parameter}
\label{sec:scatter plots}

\begin{figure*}
\includegraphics[width = \textwidth]{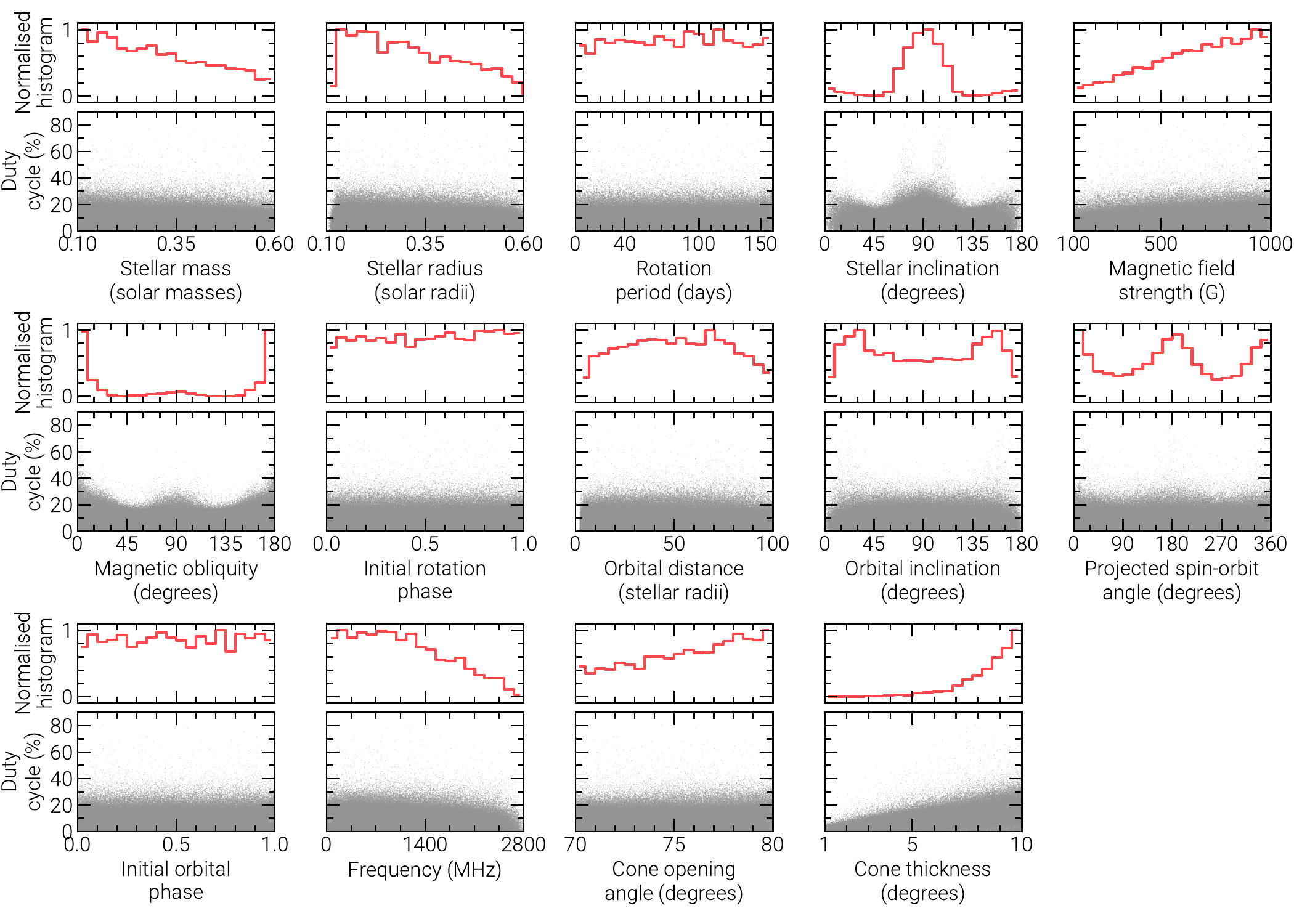}
\caption{Scatter plot of the signal visibility for each sample in the Monte Carlo simulation against each input parameter for of the model, as well as the true spin-orbit angle. The top sub-panel in each shows the normalised histogram of the number of systems over each parameter where the duty cycle exceeds 20\%. Each histogram has 20 bins, which span the ranges listed each parameter in Section~\ref{sec:samples}.}
\label{fig:scatter plots}
\end{figure*}

In Figure~\ref{fig:scatter plots}, we show the scatter plot of the duty cycle of each system in the Monte Carlo simulation performed in Section~\ref{sec:bias general} against each of the model parameters.

\bsp
\label{lastpage}
\end{document}